\documentclass[aps, pra, a4paper, showpacs, english, twocolumn, 10pt]{revtex4-1}
\usepackage{bbm, amsthm, bm,textcomp, nicefrac,geometry,ragged2e}
\usepackage[dvipsnames]{xcolor}
\usepackage{float}
\usepackage[bbgreekl]{mathbbol}
\usepackage{graphicx,epstopdf,color,verbatim,enumitem,ulem}
\usepackage{pbox,array}
\usepackage{mathrsfs}
\usepackage{amssymb}
\usepackage{stackrel}
\usepackage{amsmath}
\makeatletter
\usepackage[caption=false]{subfig}
\usepackage[T1]{fontenc}
\usepackage[caption=false]{subfig}
\usepackage{babel}
\addto\captionsenglish{}

\begin{document}
\title{Delocalized information in quantum networks}
\author{J. Miguel-Ramiro$^{1}$ and W.~D\"ur$^1$}
\affiliation{$^1$ Institut f\"ur Theoretische Physik, Universit\"at Innsbruck, Technikerstra{\ss}e 21a, 6020 Innsbruck, Austria}

\date{\today}

\begin{abstract}
We consider entanglement-based quantum networks where information is stored in a delocalized way within regions or the whole network. This offers a natural protection against failure of network nodes, loss and decoherence, and has built-in security features. Quantum information is transmitted within the network by performing local measurements on individual nodes only. Information can be localized within regions or at a specific node by collaborative actions using only entanglement within a region, or sometimes even without entanglement. We discuss several examples based on error correction stabilizer codes, Dicke states and correlation space encodings. We show how to design fully functional networks using encoded states or correlation space resources.
\end{abstract}
\maketitle

\section{Introduction}\label{Intro}
Quantum networks promise to be one of the first applications of an upcoming quantum technology \cite{Kimble_2008,Wehner2018} (see also e.g. \cite{Pirker2,Pirker1,cuquet,Perseguers1,Perseguers2,Meter2013b,VanMeter2014book,Bruss1,Bruss2,Gyongyosi2017,Caleffi17,Cacciapuoti18,Caleffi18a,Dahlberg2018,Dam_2017,Hahn2019,Pant2019,Markham2019}). They are the quantum counterpart of classical networks, where quantum information rather than classical information is distributed, stored and processed. Of particular importance is the distribution and storage of entanglement. Such entangled quantum states do not have a classical analog, and offer various applications ranging from quantum cryptography and conference key agreement to distributed metrology and distributed quantum computation \citep{Ekert,Bennett1992,Bennett84,sensing1,sensing2,sensing3,clocks,Khabiboulline_2019,Ciampini2016,Ciracdis,Pirker17}.

Usually such a quantum network is thought of consisting of multiple nodes or parties \footnote{Throughout this article, we use party, node and site synonymously to refer to a single network node that is located at and operated by a specific party.} that are connected by quantum channels \cite{Wehner2018}. Upon request, quantum information is directly distributed, or entangled states are generated on demand between different parties.
These entangled states can then be used for different applications, including the distribution of quantum information via teleportation. Alternatively a top-down approach to quantum networks is used \cite{Pirker2,Pirker1}, where entangled resource states shared among network nodes serve to guarantee the required functionality. In both cases quantum information is stored locally at a given node.

Here we analyze a different approach to quantum networks, where quantum information is stored in a delocalized way within network regions or the whole network. Such an approach offers several advantages and interesting features that we illustrate and analyze in detail:
 \begin{itemize}
 \item[(i)] Natural protection against failure of network nodes, loss and decoherence during storage and transport.
 \item[(ii)] Built-in security features, such as limited accessible information per network node.
 \item[(iii)] Encoding/Decoding and processing of information using only local operations or limited entanglement.
 \end{itemize}
 We clarify these features as follows: (i) Since information is not localized at a given network node, failure or loss of individual nodes does not destroy encoded information completely, but only disturbs it to a certain extend. The same is true for processing or transport of information. While in a standard approach, failure or loss of any of the nodes involved in a transport process results into complete information loss or loss of entanglement, in the approach we follow here the failure of one (or several) individual nodes may have only a very limited influence, thereby minimizing the influence of network node failures. (ii) With respect to security of the stored information, the accessible information per node is bounded and can be made arbitrarily small. This implies that multiple nodes need to cooperate in order to access the information, while it is protected against malicious parties. In this context it is also interesting to note that the entanglement shared between an individual party and the rest of the network can be small \cite{Gross07}, significantly less than one ebit. (iii) Since information is no longer stored in its bare form, one has to think of encoding, decoding and processing of information. Ideally, this should be done by local operations on individual nodes only, however it may also require some (restricted) amount of shared entanglement. In fact we find that processing of information, in particular transport among an entanglement-based network, is always possible using local operations only \cite{Gross072,wires}.

We consider two different scenarios: (a) Storage networks, where quantum information is stored in a distributed way among all or multiple nodes, and (b) generic networks with full functionality, including transport, that are comprised of different connected regions. Each region consists of multiple network nodes and corresponds to a single logical qubit. Regarding (a), we analyze several kind of encodings, where the logical basis states are given by codewords of error correction stabilizer codes, Dicke states \cite{Dicke1,Dicke2,Dicke3}, or resource states that have been discussed in the context of quantum computation in correlation space \cite{Gross07,Gross072,wires}. The usage of error correction codes for storage is well known and has been widely discussed. Such an approach offers protection against noise or loss, however requires active error correction. When used in a distributed scenario as we consider here, entanglement or non-local operations are required to detect and correct errors. Dicke state encodings in contrast have passive, built in protection features. Even without active error correction, quantum information is only slightly disturbed by loss, decoherence and node failures of a restricted amount of parties. Furthermore, the information that is distributed among the network can in some cases be probabilistically localized (using only local operations by all parties, or some restricted amount of entanglement). Finally, we investigate resource states for quantum computation in correlation space, particularly so-called quantum wires \cite{wires}. There quantum information is encoded in a virtual space that is not directly linked to physical qubits forming the quantum states. Local measurements on physical qubits allow one to manipulate, process and read-out information stored in correlation space. We analyze and discuss how information encoded on the correlation space is distributed throughout the physical particles. Importantly, a downloading process can be performed, i.e. the quantum information stored in a virtual space can be mapped to physical sites \cite{downloadDur}. This is however in general a probabilistic process that can be made quasi-deterministic if the download should take place to an arbitrary node within a small region. Noise and imperfections on physical qubits during storage or manipulation influence the quantum information stored in correlation space, in many cases however in a strongly reduced way.

Regarding (b), we consider multi-party entangled states comprised of logical qubits, that act as resource states for a quantum network with full functionality. Each of the logical qubits is associated with a region that consists of several parties, and we consider entangled states of multiple logical qubits shared among the network. Depending on the choice of resource state, point-to-point communication, generation of bipartite entanglement or the generation of arbitrary logical graph states between network regions or individual nodes is possible.
In the context of correlation space encodings we consider complex (multi-dimensional) resource states. Then also processing, or more concretely transport and of information among the network and download to individual sites, is of importance. We show that for all cases we consider, the transport can be done solely by {\it local} operations on individual physical nodes, so no extra entanglement is required \cite{walgate}. We also study the influence of noise and imperfections on such transport and downloading processes. For correlations space resources, we introduce the notion of transport universality. Note that the required functionality of resource states in a network differs from full scale measurement-based quantum computation as considered in \cite{Gross07}. This implies that a much larger class of resource states is suitable in a communication scenario as we consider here. We find that information transport and download can be done solely by local operations, and analyze the influence of noise and imperfections on physical sites.

The paper is organized as follows. In Sec. \ref{Backg} we describe the problem setting and introduce required terminology and notation, in particular regarding Dicke states, measurement-based quantum computation and computation in correlation space. In Sec. \ref{storage} we consider the storage scenario (a), and analyze uploading, downloading and localization of information. We study the influence of noise and imperfections for all combinations of Dicke-state encodings in Sec \ref{dickeenconding}, while we investigate various correlation space encodings in Sec. \ref{correlationencoding}. In the latter case, we determine the effective noise in correlation space, and investigate the dependence on error positions in physical space. In Sec. \ref{LogicalNetworks} and Sec. \ref{sec:Correlationtransport} we consider scenario (b), i.e. full quantum networks built from encoded (logical) states, or by combining correlation space resources respectively. 

\section{Background}\label{Backg}

\subsection{Problem setting}
We consider $N$ spatially separated parties, each holding a two-level system (qubit). The parties share a pre-defined entangled state that serves to store one (or several) logical qubits in this network in a delocalized way. Information needs to be uploaded from a given site into the network, downloaded to a specific site with the help of other parties, as well processed. In all cases, this corresponds to a collaborative process where some of the other parties perform measurements on their qubit. In the following we will review some of the underlying concepts. We start by describing measurement-based information processing, followed by the description of two different ways to encode quantum information, using Dicke states and correlation space resources. The required processes and concepts are then discussed in following sections.



\subsection{Measurement Based Quantum Computation}
Measurement Based Quantum Computation (MBQC) is an alternative model for quantum computation in opposition to quantum
logical circuits or quantum Turing machines. MBQC has no direct
classical counterpart, and is based on the implementation of single qubit
measurements on a resource state  \cite{MBQC1,MBQC2}. Any operation applied to
an arbitrary input can be reproduced by an adequate measurement pattern
on the resource state. Different classes of resource states have been
studied in the MBQC context. In particular, resources for MBQC
based on tensor networks \cite{Gross07,Gross072} are of special interest for the purposes of this
paper.

\subsection{Encoded resource states - Dicke states\label{subsec:Encoded-resource-states}}
\begin{figure}[h]
\includegraphics{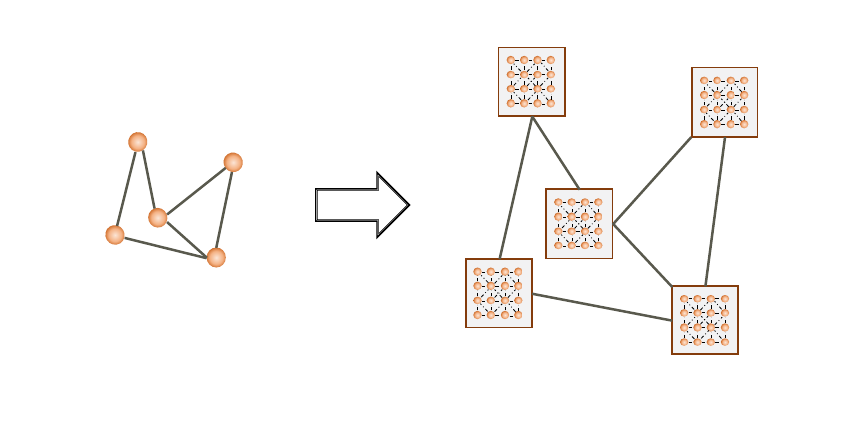}
\caption{\label{fig:encoding} Physical qubits are encoded into larger systems (logical qubits) where storage or quantum processing  is performed in a protected way.}
\end{figure}

A natural way to protect information in quantum computation consists
in distributing the information of some particular state into a larger
system of more physical particles. Given an arbitrary single qubit
state $\left|\varphi\right\rangle =\alpha\left|0\right\rangle +\beta\left|1\right\rangle $,
it can be encoded into a larger logical system of $n$ particles 
i.e. $\left|\varphi_{L}\right\rangle =\alpha\left|0_{L}\right\rangle +\beta\left|1_{L}\right\rangle$,
with some appropriate choice of $\left|0_{L}\right\rangle ,\left|1_{L}\right\rangle$ , two
orthogonal states of $\mathbb{C}^{2^{n}}.$ With a proper choice of the encoding, the information of the
state can be shielded against errors, losses or intervention of any physical
constituents of the system. Logical blocks are then entangled
(see figure \ref{fig:encoding}) and information is processed
at the logical level. In \cite{Gross07,Gross072}, computational universality of MBQC
resources is analyzed in detail, and novel resources are proposed.
However, we are not interested in computational properties of the
encoded resource states, but in their storage and communication characteristics, i.e.
how the information can be diluted throughout the parties and how one can eventually
localize it in some particular physical system.

The construction of encoded resources states 
is a passive alternative to error correction codes (where errors are identified and corrected), such that storage and processing of quantum information in a protected way is guaranteed. In this paper, we aim to find good
choices of orthogonal states $\left|0_{L}\right\rangle:=\left|\psi_{0}\right\rangle $,
$\left|1_{L}\right\rangle:=\left|\psi_{1}\right\rangle $ in terms
of robustness and strength of the encoded resource states for storage
purposes. Our analyses are restricted to the use of local operations
assisted by classical communication (LOCC) at a physical level.
For a given resource state, only operations on single qubits
are available.

Dicke states \cite{Dicke1} are an interesting class of quantum states with different applications in quantum information \cite{dicke5,dicke6,dicke7} and experimentally obtainable. 
A Dicke state of
$n$ qubits and $k$ excitations is a symmetric state of the form:

\begin{equation}
\mathscr{\mathcal{D}}_{\left|n,k\right\rangle }=\binom{n}{k}^{-\frac{1}{2}}\sum_{i}\mathcal{P}_{i}\left(\left|1\right\rangle ^{\otimes k}\left|0\right\rangle ^{\otimes n-k}\right),\label{eq:dicke}
\end{equation}
where the operator $\mathcal{P}$ represents
all the possible permutations for a given number of particles and excitations.
For instance, for $k=0$, the states reduces to the product state
$\left|0\right\rangle ^{\otimes n}$ and for $k=1$ the state is simply the
$W_{n}$ state \cite{wstate}. Note that any two Dicke states with
different number of excitations $k\neq k'$, $\mathscr{\mathcal{D}}_{\left|n,k\right\rangle }$,
$\mathscr{\mathcal{D}}_{\left|n,k'\right\rangle }$, are orthogonal
to each other.

\subsection{Correlation space\label{subsec:Correlation-space}}
Several classes of computational resources \cite{Gross07,Gross072} have been introduced
as universal resources for MBQC. These resource states are
defined within a tensor network formalism, with some particular boundary
conditions, where quantum information is processed in a virtual space
called correlation space. In particular, we are interested in qubit
computational wires introduced in \cite{wires}. A computational wire is
a family of pure states formed by a one dimensional chain of two-level
systems which fulfils two properties, i.e. it is preparable from
a product state by nearest-neighbour interaction and the entanglement
between the left and right sides of the chain approaches one ebit
in the limit of large number of qubits. Notice however, that
this does not imply that the entanglement of an individual physical particle
with the rest of the chain is one ebit.

\begin{figure}[h]
\includegraphics{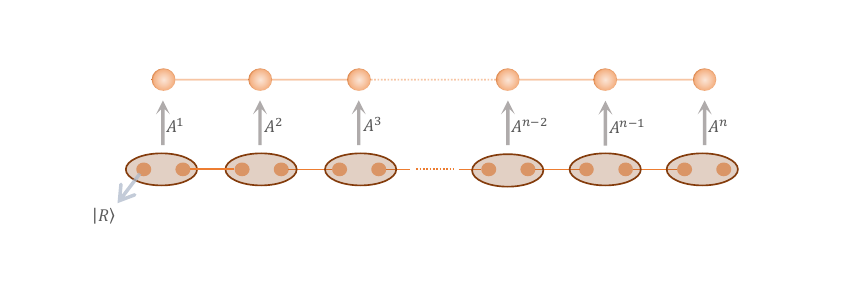}
\caption{\label{mpsform} Graphical representation of a Matrix Product State in terms of the valence-bond picture. Each physical particle is associate with two virtual systems, each of one shares a maximally entangled state with their neighbours. The projector of the virtual space into each physical system at every site is given my matrices $A[s_{i}]$.}
\end{figure}

A pure quantum state of a chain of $n$ qudits can be described by a matrix
product state (MPS) representation \cite{mps1,mps2} with e.g. open-boundary
conditions (see figure \ref{mpsform}),
\begin{equation}
\Phi\left(\left|L\right\rangle \right)_{1}^{n}=\underset{s_{i}=0}{\sum^{d-1}}\left\langle R\right|A\left[s_{n}\right]\cdots A\left[s_{1}\right]\left|L\right\rangle \left|s_{1}\cdots s_{n}\right\rangle ,\label{eq:mpsbasic}
\end{equation}
where $s_{i}=\left\{ 0,\ldots,d-1\right\}$. Quantum information
is processed in a virtual $D$ dimensional vector space (correlation
space) where the $D\times D$
matrices $A\left[i\right]$ live, as well as  the $D$ dimensional vectors $\left|L\right\rangle ,\left|R\right\rangle$,
which represent the left and right boundary conditions. The MPS is denoted as $\Phi\left(\left|L\right\rangle \right)_{1}^{n}$
since one can consider that the correlation space is in the state $\left|L\right\rangle $ \cite{Gross07}.
Any measurement on a single physical qubit is translated into some
operation acting on the correlation space state.

\begin{figure}[h]
\includegraphics{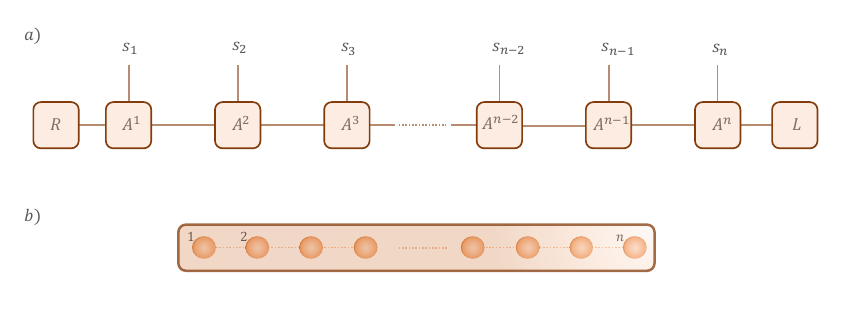}
\caption{\label{mpsform2}a) Graphical representation of a matrix product state in terms of tensor networks. Every block is a rank-3 tensor, and open indices are the physical indices (see \cite{Gross07,mps2} for a detailed explanation of graphical rules for tensor network).  b) Simplified illustration of a quantum computational wire, where qubits are attached in a 1D wire with arbitrary local entanglement and non-vanishing two point correlation functions.}
\end{figure}

One particular instance of 1D structure described within this formalism, and denoted as quantum computational wire  \cite{wires}, has been shown to be an appropriate building block for constructing resources universal for quantum
computation. For this kind of wires, it
is shown \cite{wires} that, by single qubit measurements, one can prepare
the correlation state in any arbitrary state $\Phi\left(\left|\varphi\right\rangle \right)_{k}^{n}$.
A local measurement on a physical particle of the wire with outcome
$\left|m_{i}\right\rangle $ is associated with the application of
the operator $A[m_{i}]$ on the correlation space. It is also known
how the information of the correlation space can be localized \cite{downloadDur} in one physical qubit and how to
upload it from a physical site \cite{MorimaeUpload}. We review these processes in Sec. \ref{correlationencoding}.
We are however interested in the communication properties
of this kind of wires under the only assistance of local measurements,
local operations and classical communication (LOCC). In this paper
we restrict ourselves to the qubit case, and where $d=D=2$.
We focus on one particular computational wire with non-vanishing
two point correlation functions and fixed boundary conditions, that we denote
as period wire (see figure \ref{mpsform2}):

\begin{equation}
\left|\psi\right\rangle =\underset{s_{i}=0}{\sum^{1}}\left\langle s_{n}\right|A\left[s_{n-1}\right]\cdots A\left[s_{1}\right]\left|+\right\rangle \left|s_{1}\cdots s_{n}\right\rangle ,\label{eq: mpswire}
\end{equation}
\begin{equation}
A[+]=\frac{1}{\sqrt{2}}G,\,\,\,\,\,\,\,\,A[-]=\frac{1}{\sqrt{2}}GT\left(\phi\right),\label{eq:matrixmps}
\end{equation}
where $G=\exp\left(i\pi X/\tau\right)$ and $T=diag\left(e^{\frac{-i\phi}{2}},e^{\frac{i\phi}{2}}\right)$.
We denote $\tau$ and $\phi$ as period and entanglement factor respectively.
Any matrix $A$ can be expressed in an arbitrary basis $\left\{ m_{0},m_{1}\right\} $
as $A[m_{i}]=\left\langle m_{i}\left|0\right\rangle \right.A[0]+\left\langle m_{i}\left|1\right\rangle \right.A[1]$.
We interpret this wire in two different ways. On the one hand, one
can think about wires as playing the role of logical blocks of qubits
as seen before, with codewords $\left|0_{L}\right\rangle :=\Phi\left(\left|\varphi_{0}\right\rangle \right)_{1}^{n},$
$\left|1_{L}\right\rangle :=\Phi\left(\left|\varphi_{1}\right\rangle \right)_{1}^{n}$,
where $\varphi_{0}$ and $\varphi_{1}$ are two orthogonal states
of the correlation space. In this case, one logical qubit is stored
in a distributed way among the wire. We study certain features
of these wires, such as entanglement of individual physical qubits,
influence of noise and losses, error propagation and information localization.

One may also interpret these wires in an alternative way. A single
wire can be conceived as a building block of a whole communication network where information
is stored in a delocalized way and transported upon request, such that direct communication between any different nodes or regions can be established. One can think about the correlation space in this scenario
as a big \textit{holographic} resource where the information is delocalized
among all the nodes, and therefore, protected.


\section{Storage of Quantum Information}\label{storage}
An essential feature of a quantum network in a communication scenario
is the storage of quantum information. In this section we explore the required
characteristics of a resource state such that information can be stored in a non-local way. We analyze families of
states that exhibit different properties. We study how errors affect the stored information, and how the information can be subsequently localized into some particular node or region, with only the assistance of LOCC.

\subsection{Processes and properties}
We start be describing the required processes, namely uploading and downloading of quantum information, and the features we analyze. This includes local entanglement, robustness against errors and loss as well as security features.

\subsubsection{Uploading\label{subsec:Uploading}}
Given a quantum state containing some information, one aims to store and protect that information against noise
or external interference. In our scenario, this is accomplished by
encoding the state into a larger network consisting of multiple network nodes or parties, such that the
information is delocalized or diluted over the network. This delocalization
process is called \textit{uploading}. Two different uploading procedures
should be distinguished.

\begin{figure}[h]
\includegraphics{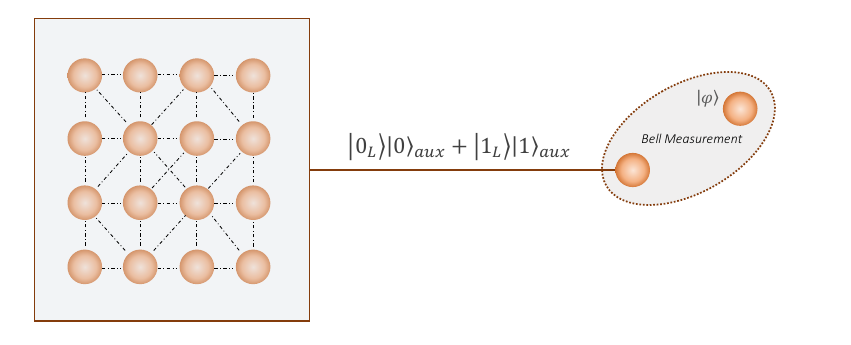}
\caption{\label{fig:Uploading}Uploading process. A Bell measurement is performed in the auxiliary system (right) to upload the information of some state into the logical level. The same picture also holds if one substitutes the logical block by a quantum wire of the form of Fig. \ref{mpsform2}}
\end{figure}

First, we consider the case where an arbitrary state
is directly encoded into a larger system of $n$ particles.
This task involves global operations (at the logical level) and can
be accomplished by constructing an entangled state between a logical
block and an auxiliary system $aux$ (see figure
\ref{fig:Uploading}). Consider the state:
\begin{equation}
\left|\psi\right\rangle =\frac{1}{\sqrt{2}}\left(\left|0\right\rangle _{aux}\left|0_{L}\right\rangle +\left|1\right\rangle _{aux}\left|1_{L}\right\rangle \right),\label{eq:uploadingstate}
\end{equation}
where the logical qubits are defined by some orthogonal codewords $\left|0_{L}\right\rangle, \left|1_{L}\right\rangle$.
This means that the system $aux$ is maximally entangled with the whole
logical block of qubits. Consider now an arbitrary state
$\left|\phi\right\rangle =\alpha\left|0\right\rangle +\beta\left|1\right\rangle$
that we want to upload into the logical level. Hence, by performing
a Bell measurement between the $aux$ system and $\left|\phi\right\rangle$,
the uploading into the logical level is deterministically achieved, i.e. the global state
of the remaining particles is now $\alpha\left|0_{L}\right\rangle +\beta\left|1_{L}\right\rangle $
(up to unitary Pauli correction operations at the logical level). We remark that the logical Pauli correction operation can in general not be implemented locally, however are irrelevant in the sense that they actually do not need to be implemented as the logical subspace is unchanged. It suffices to relabel the basis states, and adapt the information processing and subsequent downloading processes accordingly.

On the other hand, the uploading process can also be done from a single node within the network without using an entangled state. In this case, only the network state (e.g. a computational wire) is used, and a single party (or possibly also several parties) aim to upload quantum information that is given in the form of an auxiliary quantum state to the logical level, thereby delocalizing the quantum information.


\subsubsection{Downloading}
The reverse process to the uploading of information is the downloading,
i.e. once information is spread out among the nodes of the network,
we require the possibility of localizing this information in one particular
physical system (again, by LOCC). In general, this is achieved by
suitable measurements on the rest of the particles. However, crucial
differences exist when localizing information with standard encoded
states (Dicke states) or in the correlation space framework. Encoding
properties defines the characteristics of the localization process. In
particular, two features are specially relevant. The ideal situation is the one where the localization
is performed in a deterministic way, and the information is downloaded in an a-priori chosen place. However,
due to the properties of each encoding state, one can be restricted to the case where only probabilistic,
or open-destination downloading process is possible. Probabilistic downloading means that information can only be localized probabilistically in a heralded way. This may be unsatisfactory in many cases. Open-destination downloading in turn is still deterministic, however the precise location can not be determined a priori. This corresponds to a probabilistic download attempt to a given site, which however can be repeated if not successful. In this way quantum information can be downloaded quasi-determistically to individual sites within a given region of small size.

We remark that for standard encodings, there are examples where the probability to localize information from two logical blocks is arbitrarily small, even for the open-destination scenario \cite{Gross07}. For correlation space encodings, this problem does not occure as we show below. The main difference is that for correlation space encodings, only a fraction of the qubits need to be measured, and unsuccessful downloaded can be repeated. This leads to a success probability approaching one to download information to an unspecified site within a region of fixed size (see Sec. \ref{correlationencoding}).

\begin{figure}[h]
\includegraphics{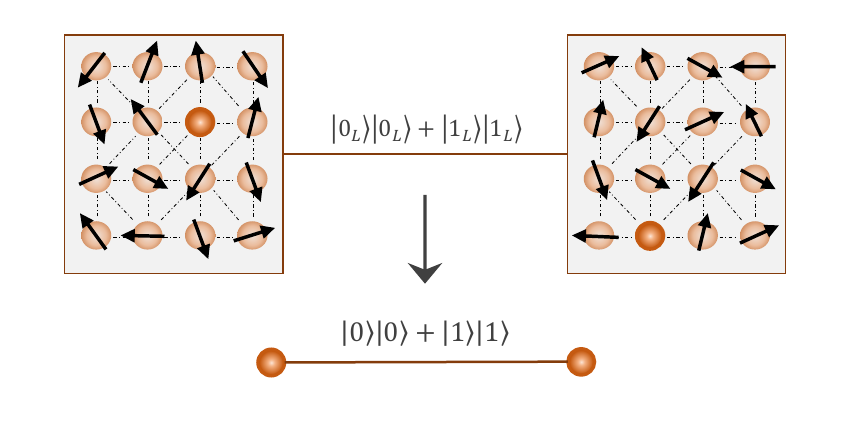}

\caption{\label{fig:Downloading} Downloading of information. Information can be localized into specific physical sites by measuring the rest of qubits in suitable basis. This process differs when localizing information of the correlation space, where one does not need to measure all the qubits of the wire, which remains functional with the unmeasured ones (see Sec.  \ref{correlationencoding})}
\end{figure}

\subsubsection{Robustness}
The effect of local errors can affect the global state of the resource, jeopardizing the stored information.
In particular, we are interested in errors occurring in single particles
when the information is delocalized, and how robust is the encoded state
under that noise. The effect of local noise
should have limited influence on the global state, in order to consider
an encoded state robust for storage. We find effective noise maps for this process. Besides, we study the stability of the
resource states under loss of particles.

In order to analyze the effective noise when errors affect particles at the physical level, one can study the following scenario. Consider the maximally entangled state
\begin{equation}
\left|\Phi^{+}\right\rangle=\frac{1}{\sqrt{2}}\left(\left|0\right\rangle _{aux}\left|0_{L}\right\rangle +\left|1\right\rangle _{aux}\left|1_{L}\right\rangle \right),\label{eq:maxentstate}
\end{equation}
where an auxiliary single particle is defined on $\mathcal{H}_{aux}=\mathbb{C}^{2}$,
entangled with a logical encoded block which lives on $\mathcal{H}_{L}=\mathbb{C}^{2^{\otimes n}}$.
The state resulting from the introduction of local noise on the physical
particles of a logical block is defined as the Choi-Jamiolkowski state \cite{Jamiokowski1972,Frowis13}
i.e.
\begin{equation}
\varGamma=id_{\mathcal{H}_{aux}}\otimes\xi^{\otimes j}\left(\left|\Phi^{+}\right\rangle \left\langle \Phi^{+}\right|\right),\label{eq:choistate}
\end{equation}
where the noisy channel $\xi^{\otimes j}$ affects $j$ particles
of the logical qubit. If one then aims to localize (by local operations)
the information of the block into one physical system $q$, it is
relevant to study how close this final state is from a perfect Bell pair
$\left|\Phi\right\rangle =\frac{1}{\sqrt{2}}\left(\left|0\right\rangle _{aux}\left|0\right\rangle _{q}+\left|1\right\rangle _{aux}\left|1\right\rangle _{q}\right),$
which would be the resulting state of a perfect localization.
Notice that $\varGamma$ contains the full information about the effective noise map, which is in principle accessible. We remark that we use the effect of the performed processes on part of a maximally entangled state only as a tool to analyze the resulting effective noise maps or the resulting process fidelity.

Similarly, the stability under losses can be analyzed in the following way. We
consider an initial state of the form $\left|\Phi^{+}\right\rangle \left\langle \Phi^{+}\right|$ (\ref{eq:maxentstate}) shared by the auxiliary particle and the logical qubit,
and we assume that some particles in the logical block are
lost. The effect of a lost particle is described by tracing out that
particle, i.e. $\varGamma_{f}=Tr_{i}\left(\left|\Phi^{+}\right\rangle \left\langle \Phi^{+}\right|\right)$,
where $i$ are the lost particles. Tracing out is equivalent to performing an
average over all the possibles outcomes of a measurement of the system
$i$ in any basis.

Finally, we compute in both cases (errors and losses) the fidelity of the final state
with respect to the initial one (\ref{eq:maxentstate}), denoted as Choi-Jamiolkowski fidelity
(CJ fidelity):
\begin{equation}
F_{CJ}=\left\langle \Phi^{+}\right|\varGamma_{f}\left|\Phi^{+}\right\rangle .\label{eq:CJfidelity}
\end{equation}
This parameter is a suitable measure to analyze how stable is a
resource encoded state against loss of particles \cite{Frowis13}. 
Note that in order to compute the CJ fidelity, we replace the state of each lost particle by an identity operator, i.e. consider a completely depolarizing map acting on each of these qubits.

\subsubsection{Local entanglement}
An essential property of the encoded states is the local entanglement
that each physical particle shares with respect to the rest. We study this characteristic, which determines the behavior,
strength and suitability of the resource states. Consider a maximally
entangled state of logical qubits (between two blocks of particles):
\begin{equation}
\left|\Phi_{L}^{+}\right\rangle =\frac{1}{\sqrt{2}}\left(\left|0_{L}\right\rangle \left|0_{L}\right\rangle +\left|1_{L}\right\rangle \left|1_{L}\right\rangle \right).\label{eq:logicalbell}
\end{equation}
We study the entropy of entanglement of a single qubit $i$ within a block by computing the Von Neumann entropy of its reduced density operator, i.e.
\begin{equation}
S(\rho_{i})=-Tr\left(\rho_{i}\log\rho_{i}\right).\label{eq: von neuman}
\end{equation}
Even though the entanglement between blocks of
logical qubits is maximal, in general this is not the case for the local entanglement of a single party. In fact, encoded states with low local entanglement are of particular importance, since this leads to advantages in order to construct and maintain the states.

Other encoding codewords can be explored based on the ones studied here. We consider some additional examples in Appendix A.

\subsubsection{Security analysis} \label{security}
Once the information is stored in the resources in a delocalized way, it is important to analyze the security of the storage, i.e. how much influence a party has about the stored state. At this point, we can make a distinction between trusted and untrusted parties. The first case analyzes how errors or losses of the physical qubits can affect the global state, and has been studied in previous sections. We focus here in the case the parties are not trusted and we hence study how much information a single party (or a group of them) can access from a encoded state, i.e. how secure is the storage against attack of parties that aim to access the information even though not authorized to do so. In order to study this, we perform a similar analysis as was done in \cite{Bell_2014} in the context of quantum secret sharing. As in \cite{Bell_2014} we use the mutual information as figure of merit. We only consider the situation where the initial state is the desired one, i.e. pure and unaltered. We remark that this does not correspond to full security analysis, but rather illustrates how much information an untrusted party can access in the network given that the resource states were distributed as they should.
Given the scenario of Sec. \ref{subsec:Uploading}, where the $q$ system can upload or encode any state, we analyze the mutual information between any party of the resource and the $q$ system. The mutual information determines how much information a particle $t$ (or a set of them) of the resource can obtain about the system $q$, and therefore, about the global encoded state once the party $q$ applies the Bell measurement (to achieve uploading), local in its system.

The quantum mutual information between two systems is defined as
\begin{equation}
I\left(q;t\right)=S\left(\rho_{q}\right)+S\left(\rho_{t}\right)-S\left(\rho_{q,t}\right),
\end{equation}
where $S$ represents the Von Neumann entropy (\ref{eq: von neuman}) of the reduced density operator of system $q$, system $t$, and both systems, respectively.
We will use the mutual information to assess information accessible by individual parties, and to determine the corresponding security features.

\subsection{Error correction codes - stabilizer encodings}
We start by briefly considering a standard encoding using codewords corresponding to error correction stabilizer codes  \cite{Graph_states}. The codewords are graph states with $|0_L\rangle = |G\rangle$ and $|1_L\rangle=\sigma_z^{\otimes N}|G\rangle$, where $|G\rangle=\prod_{(k,l) \in E} U_{kl}|+\rangle^{\otimes N}$. Here $U_{kl}={\rm diag}([1,1,1,-1])$ is a phase gate, $|+\rangle=(|0\rangle+|1\rangle)/\sqrt{2}$ and the graph $G$ is specified by a set of edges $E$. The error correction properties of such a code depend on the choice of $G$, and codes to protect against a fixed number of arbitrary single-qubit errors can be designed.

In order to make use of the error correction features, active error correction needs to be performed. However, decoding and syndrome readout require global access to the state, i.e. additional entanglement. In contrast, the downloading process to any specific site is deterministic, and can be accomplished by simply measuring all other qubits in the $Z$-basis. Notice however that in this way the error correction features of the code are not employed. The entanglement of any qubit with the rest of the system is maximal (for any connected graph).

\subsection{Dicke states}\label{dickeenconding}
We continue our study by considering Dicke state encodings. 
We assume codewords of the form
\begin{equation}
\left|0_{L}\right\rangle =\left|n,k_{1}\right\rangle ,\left|1_{L}\right\rangle =\left|n,k_{2}\right\rangle ,\label{eq:codewo}
\end{equation}
where $\left|n,k\right\rangle=\mathscr{\mathcal{D}}_{\left|n,k\right\rangle}$
is an abbreviate notation representing a Dicke state of $n$ qubits
and $k$ excitations (\ref{eq:dicke}).

\subsubsection{Uploading}
A general uploading process can be accomplished from a state of the
form (\ref{eq:uploadingstate}) by proceeding as explained in section \ref{subsec:Uploading}.

\subsubsection{Downloading}
In order to analyze how information can be localized from the logical level into one physical constituent, we consider a downloading process (with LOCC) from two entangled blocks (\ref{eq:logicalbell}) where information is eventually localized into one of the physical qubits
of each block (sites $k$ and $k'$), such that we end up
with a Bell pair between the two final qubits (see figure
\ref{fig:Downloading}). Essentially, this procedure is accomplished
by performing suitable measurements in all except one particle in each block. This
process can only succeed deterministically for an a-priori fixed site if the
local entanglement of the logical qubit is maximal.

Consider an initial logical Bell state of the form (\ref{eq:logicalbell}) and two fixed sites $k$ and $k'$. The initial entanglement
at the logical level is maximal, but the local entropy of entanglement
for a single site $k$ is $E_{inital}\left(\rho_{k}\right)\leq1.$
If downloading succeeds deterministically, information is localized in the sites $k$ and $k'$,
and the final state is a perfect Bell pair. Therefore,
the entropy of entanglement of particle $k$ at the end is $E_{final}\left(\rho_{k}\right)=1.$
Since the process only involves LOCC and single-site
measurements, one can directly conclude that localization only succeeds
deterministically when the local entanglement of the particle
in the initial state is maximal.

If the local entanglement is not maximal, one has to face with a probabilistic
localization (in the sense that the final place where the information
is localized cannot fixed a priori), or a deterministic but imperfect
(in terms of the fidelity of the final state) downloading.

In case the local entropy of entanglement is maximal,
consider e.g. $\left|0_{L}\right\rangle =\left|0\right\rangle ^{\otimes n},\left|1_{L}\right\rangle =\left|1\right\rangle ^{\otimes n}$,
localization is simply accomplished by performing $X$ measurements
on all except one particle of each logical block. This is in fact also the case for any graph state encoding, where $Z$ measurements are enough to localize information.


\subsubsection{Robustness under errors}
\begin{figure*}[ht!]
\subfloat{\includegraphics[height=1.8in]{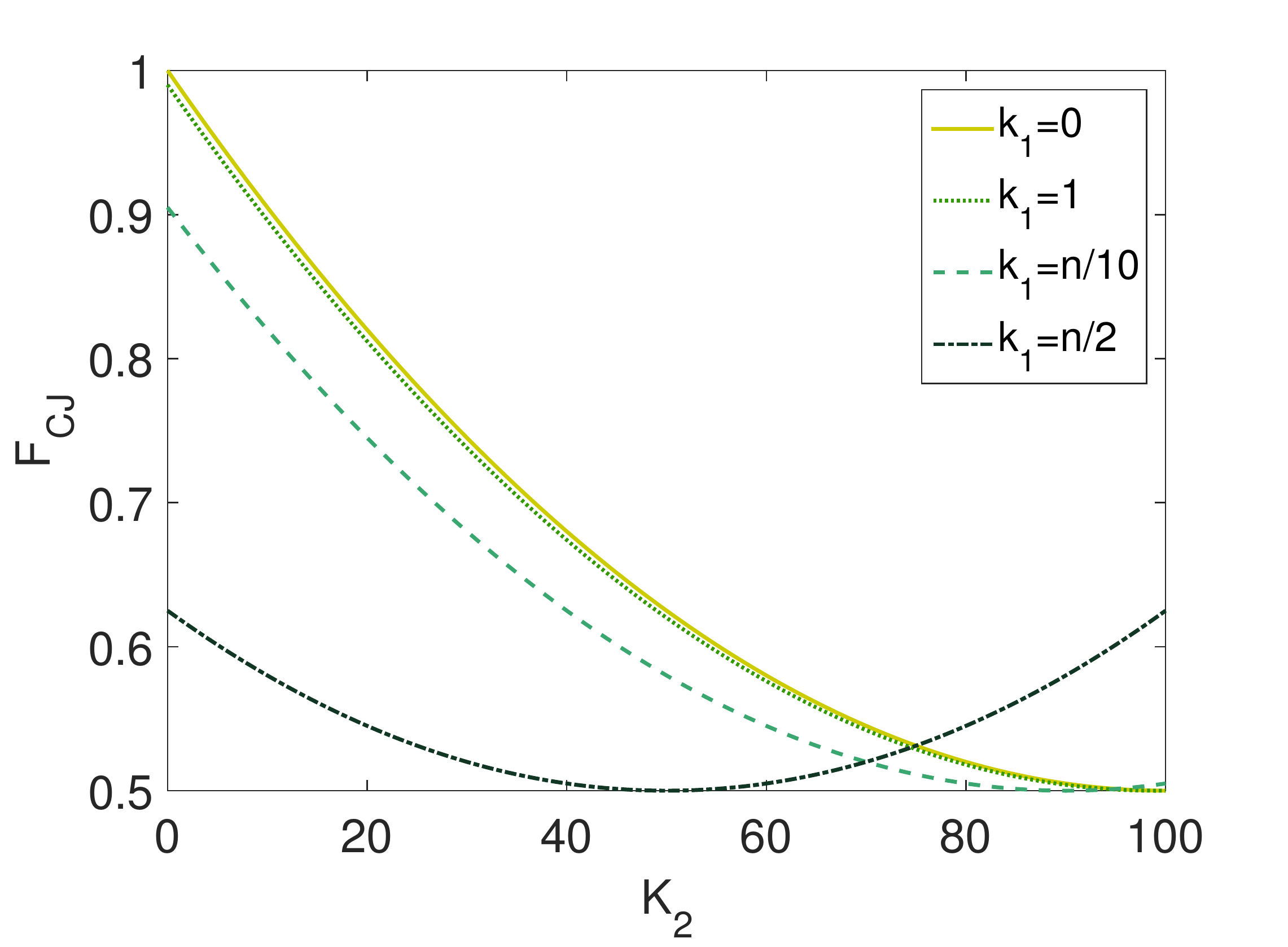}}
\subfloat{\includegraphics[height=1.8in]{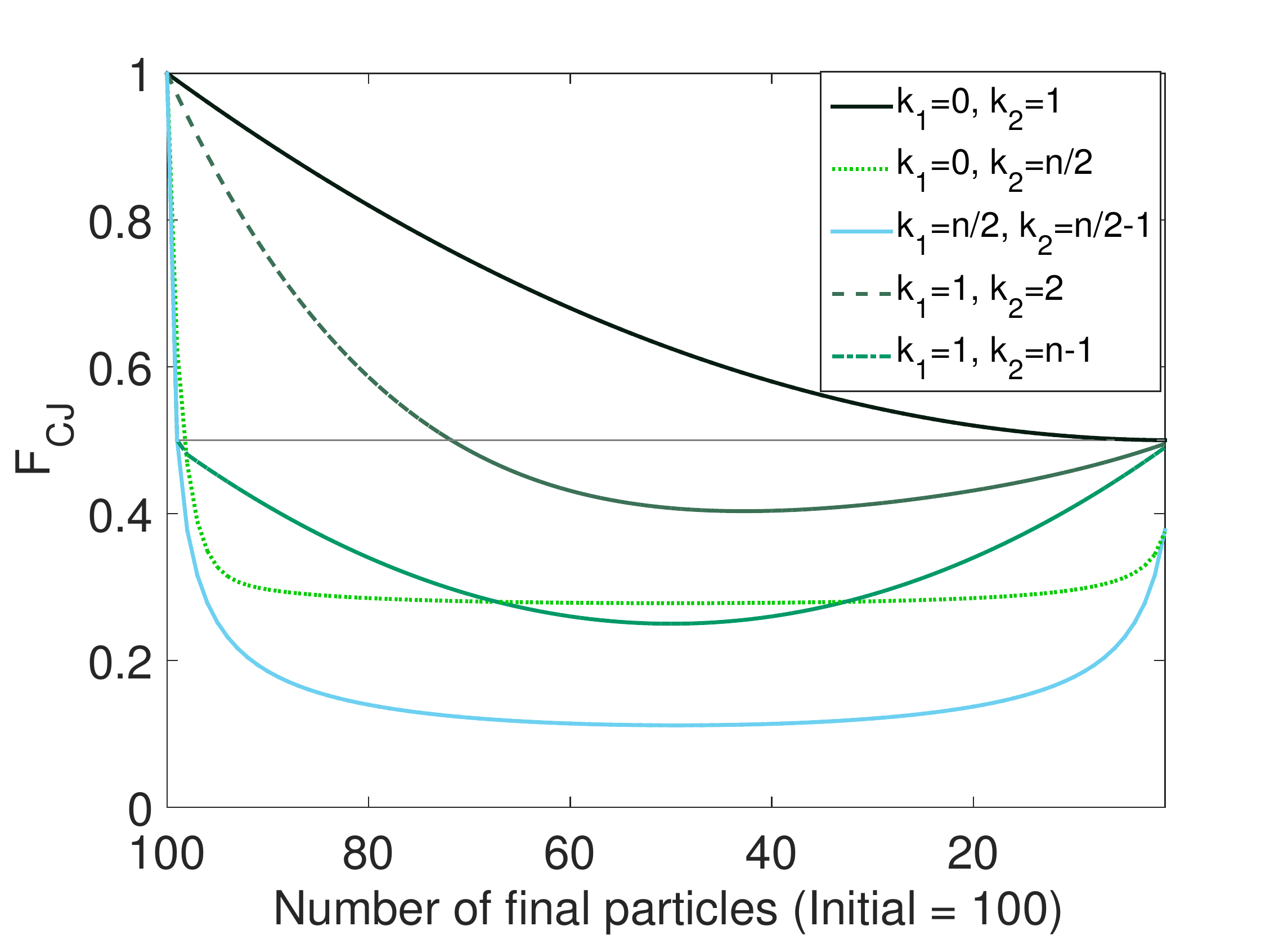}}
\subfloat{\includegraphics[height=1.8in]{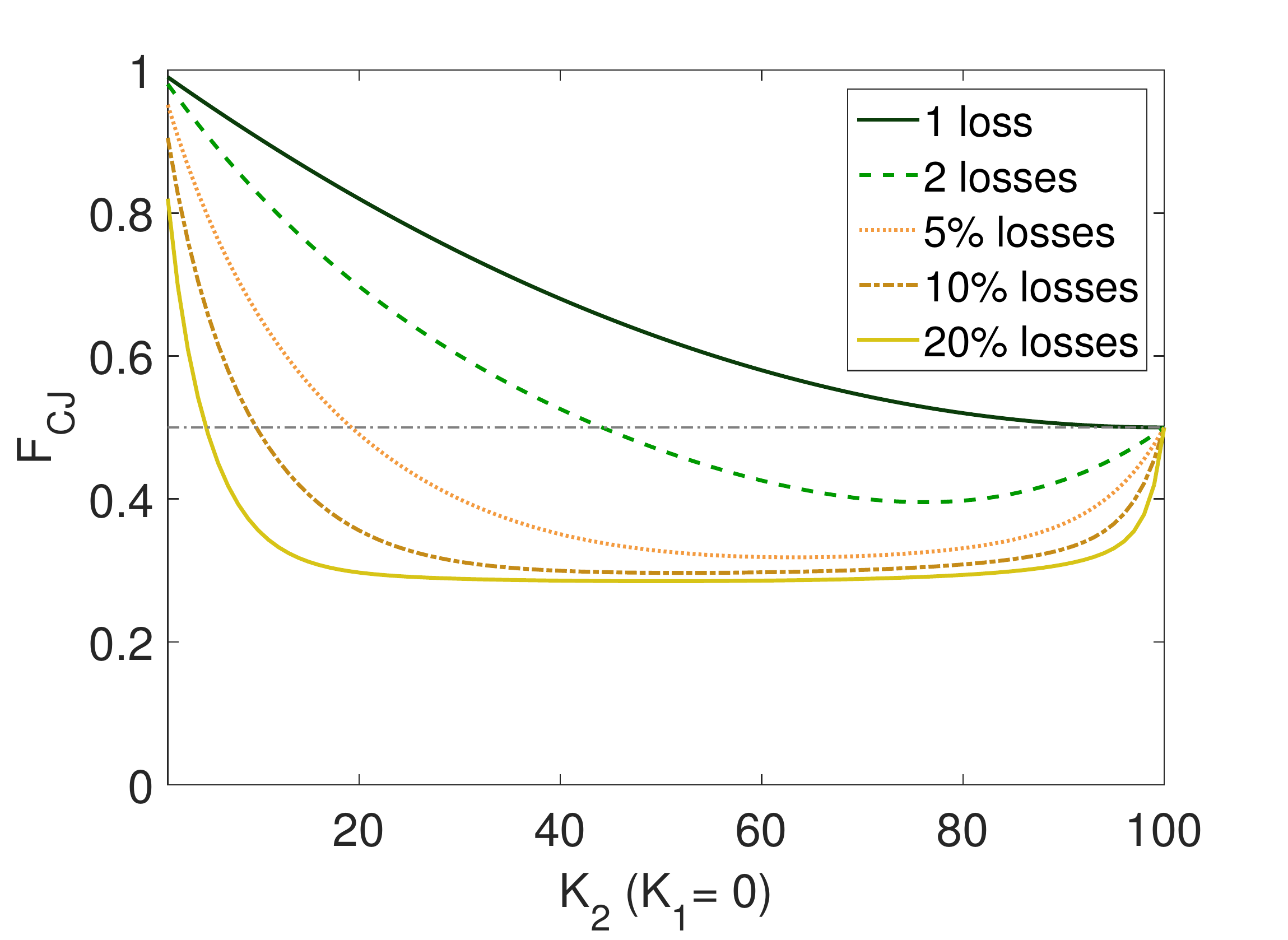}}
\caption{\label{fig:Dickeloss} (a) CJ fidelity for one loss and for one codeword Dicke state fixed as a function of the other Dicke state (note scaling of Y axis). (b) CJ fidelity for given codewords as a function of the number of lost qubits. (c) CJ fidelity for the codewords  $\left|0_{L}\right\rangle =\left|n,0\right\rangle$ fixed and different number of losses.}
\end{figure*}

In order to compute the CJ fidelity (\ref{eq:CJfidelity}) we make
use of some properties of Dicke states. Any Dicke state (\ref{eq:dicke})
of $n$ particles and $k$ excitations can be $j$-times decomposed
as \cite{Dicke3}:

\begin{equation}
\left|n,k\right\rangle =\left[\frac{n!}{\binom{n}{k}\binom{n}{j}}\right]^{\frac{1}{2}}\stackrel[q=q^{'}]{q^{''}}{\sum}\frac{\left|j,q\right\rangle \left|(n-j),(k-q)\right\rangle }{\sqrt{q!(j-q)!(k-q)!(n-k-j+q)!}},\label{eq:dickedecompos}
\end{equation}
where $q'=\max\left(0,(j+k-n)\right)$ and $q''=\min\left(j,k\right)$,
for $q''>q'.$
On the other hand, given a state $\rho_{n,k}=\left|n,k\right\rangle \left\langle n,k\right|$,
tracing out $m$ number of particles leads to:
\begin{equation}
Tr_{m}\left(\rho_{n,k}\right)=\stackrel[i=0]{k}{\sum}\frac{\binom{n-n_{f}}{i}\binom{n_{f}}{k-i}}{\binom{n}{k}}\rho_{n_{f},k-i},\label{eq:traceout1}
\end{equation}
where $n_{f}$ is the final number of particles, i.e. $m=n-n_{f}$. In particular, for any element $\left|n,k_{1}\right\rangle \left\langle n,k_{2}\right|$
we find:
\begin{multline}
Tr_{m}\left(\left|n,k_{1}\right\rangle \left\langle n,k_{2}\right|\right)= \\
\stackrel[i=0]{\min(k_{1},k_{2})}{\sum}\frac{\binom{n-n_{f}}{i}\sqrt{\binom{n_{f}}{k_{1}-i}\binom{n_{f}}{k_{2}-i}}}{\sqrt{\binom{n}{k_{1}}\binom{n}{k_{2}}}}\left|n_{f},\left(k_{1}-i\right)\right\rangle \left\langle n_{f},\left(k_{2}-i\right)\right|.\label{eq:traceout2}
\end{multline}
Computing the CJ fidelity (\ref{eq:CJfidelity}) is hence straightforward for any pair of codewords $\left|0_{L}\right\rangle =\left|n,k_{1}\right\rangle ,\left|1_{L}\right\rangle =\left|n,k_{2}\right\rangle $
and for any number of losses:
\begin{widetext}
\begin{equation}
F_{CJ}=\stackrel[i,j=1]{2}{\sum}\left(\mathcal{C}\stackrel[q=q']{q''}{\sum}\binom{n-n_{f}}{q}\left[\frac{\binom{n_{f}}{k_{i}-q}\binom{n_{f}}{k_{j}-q}}{\binom{n}{k_{i}}\binom{n}{k_{j}}q!(n-n_{f}-q)!(k_{i}-q)!(k_{j}-q)!(n_{f}-k_{i}+q)!(n_{f}-k_{j}+q)!}\right]^{\frac{1}{2}}\right),\label{eq:CJfid}
\end{equation}
\end{widetext}
where
\begin{equation}
\mathcal{C}=\frac{n!}{4\binom{n}{n-n_{f}}\sqrt{\binom{n}{k_{i}}\binom{n}{k_{j}}}},\label{eq:}
\end{equation}
and with $q'=\max\left[0,(\max\left(k_{1},k_{2}\right)-n_{f})\right]$ and
$q''=\min\left[\left(n-n_{f}\right),\min\left(k_{1},k_{2}\right)\right]$.
In figure \ref{fig:Dickeloss} one can see an illustrative analysis
of the CJ fidelity for different codewords configurations. Two
main conclusions can be extracted. On the one hand, one obtains a higher CJ fidelity
when the codewords are close to each other in terms of number of excitations. Moreover, if both Dicke states are close to the limit of zero or maximal number of excitations, the CJ fidelity is again
increased. Therefore, it turns out that the best choice of Dicke state
codewords, if one is only interested on the robustness under losses,
is $\left|0_{L}\right\rangle =\left|n,0\right\rangle ,\left|1_{L}\right\rangle =\left|n,1\right\rangle $,
or the symmetric choice: $\left|0_{L}\right\rangle =\left|n,n\right\rangle ,\left|1_{L}\right\rangle =\left|n,n-1\right\rangle $.

However, we find that a conflict of interest exists if one requires other properties more than just robustness
under losses. For instance, as shown above, the success probability of localizing information increases with the local entanglement.

\subsubsection{Local entanglement {\label{subsec:Local-entanglement-(Entropy}}}
We analyze now the entanglement properties of the Dicke-type encoding.
Consider a logical maximally entangled state of the form (\ref{eq:logicalbell}).
We study the local entanglement that a physical particle shares with
the rest of the system by computing the Von Neumann entropy (\ref{eq: von neuman}) of its reduced density
matrix. For the encoding codewords $\left|0_{L}\right\rangle =\left|n,k_{1}\right\rangle ,\left|1_{L}\right\rangle =\left|n,k_{2}\right\rangle $,
and from the properties introduced before, it is easy to see that
the entropy of entanglement of a single particle is:
\begin{multline}
S\left(\rho_{A}\right)=-\left(\stackrel[i=1]{2}{\sum}\frac{n-k_{i}}{2n}\right)\log\left(\stackrel[i=1]{2}{\sum}\frac{n-k_{i}}{2n}\right)- \\
\left(\stackrel[j=1]{2}{\sum}\frac{k_{j}}{2n}\right)\log\left(\stackrel[i=1]{2}{\sum}\frac{k_{j}}{2n}\right).\label{eq:entropydicke}
\end{multline}

\begin{figure}[h]
\includegraphics[height=2.3in]{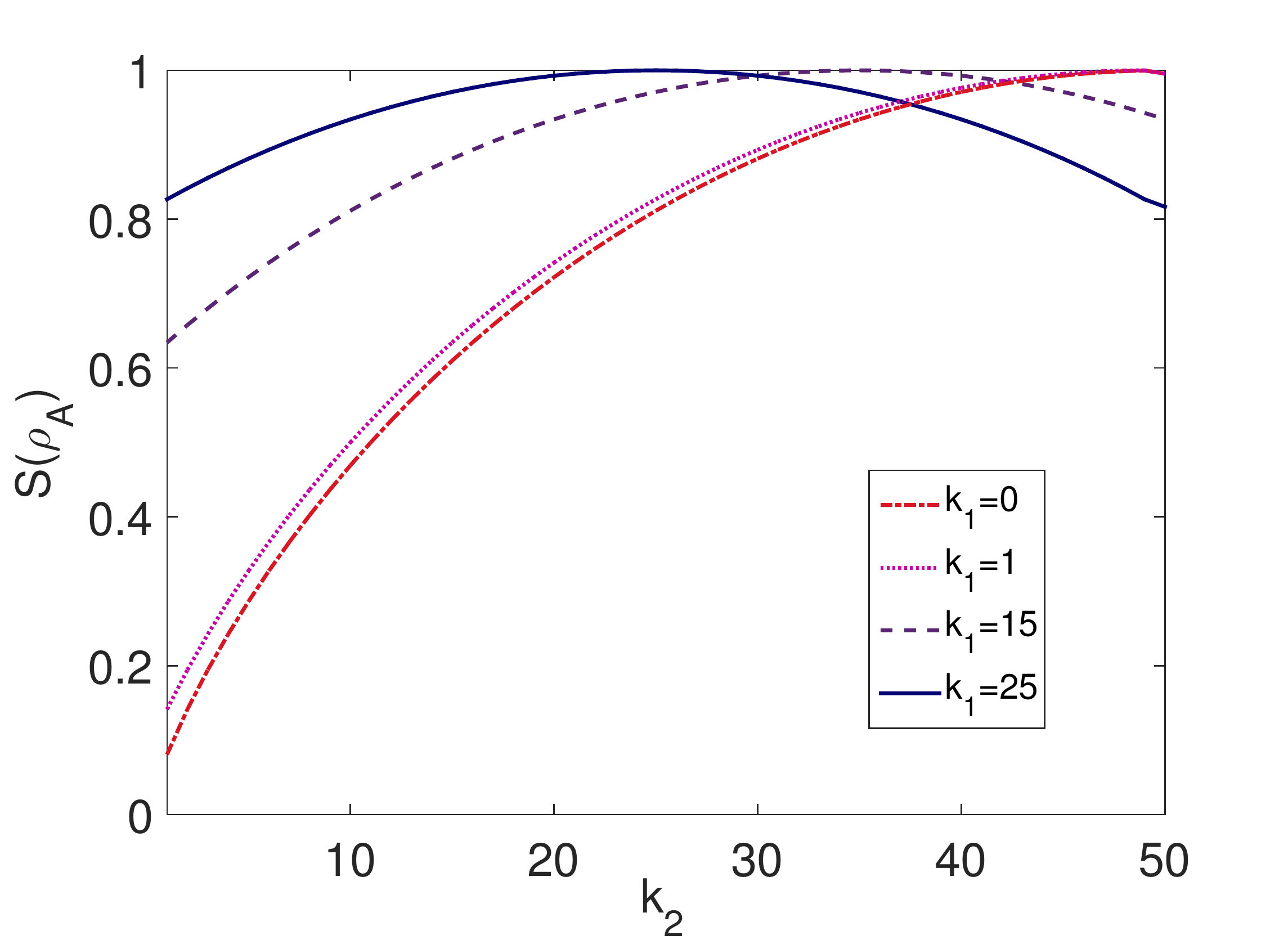}

\caption{\label{fig:entropy1} Local entropy of
entanglement of a single particle (\ref{eq:entropydicke}) for different codewords configurations.}
\end{figure}

The entropy of entanglement is minimal (see figure \ref{fig:entropy1})
for encoding codewords which are, again, close to each other in terms
of excitations and close to the boundaries with small or large (due
symmetry) number of excitations. This result is consistent with the
analysis of the robustness against losses, i.e. the less local entanglement,
the more delocalized the information is and the more stability against
losses.

An explanatory example is the extreme case with codewords $\left|0_{L}\right\rangle =\left|0\right\rangle ^{\otimes n},\left|1_{L}\right\rangle =\left|1\right\rangle ^{\otimes n}$,
where the local entanglement is maximal and the stability is minimal,
i.e. when a single particle is lost, the information and entanglement
is completely destroyed.

\subsubsection{Security analysis}
\begin{figure}
\includegraphics[height=2.3in]{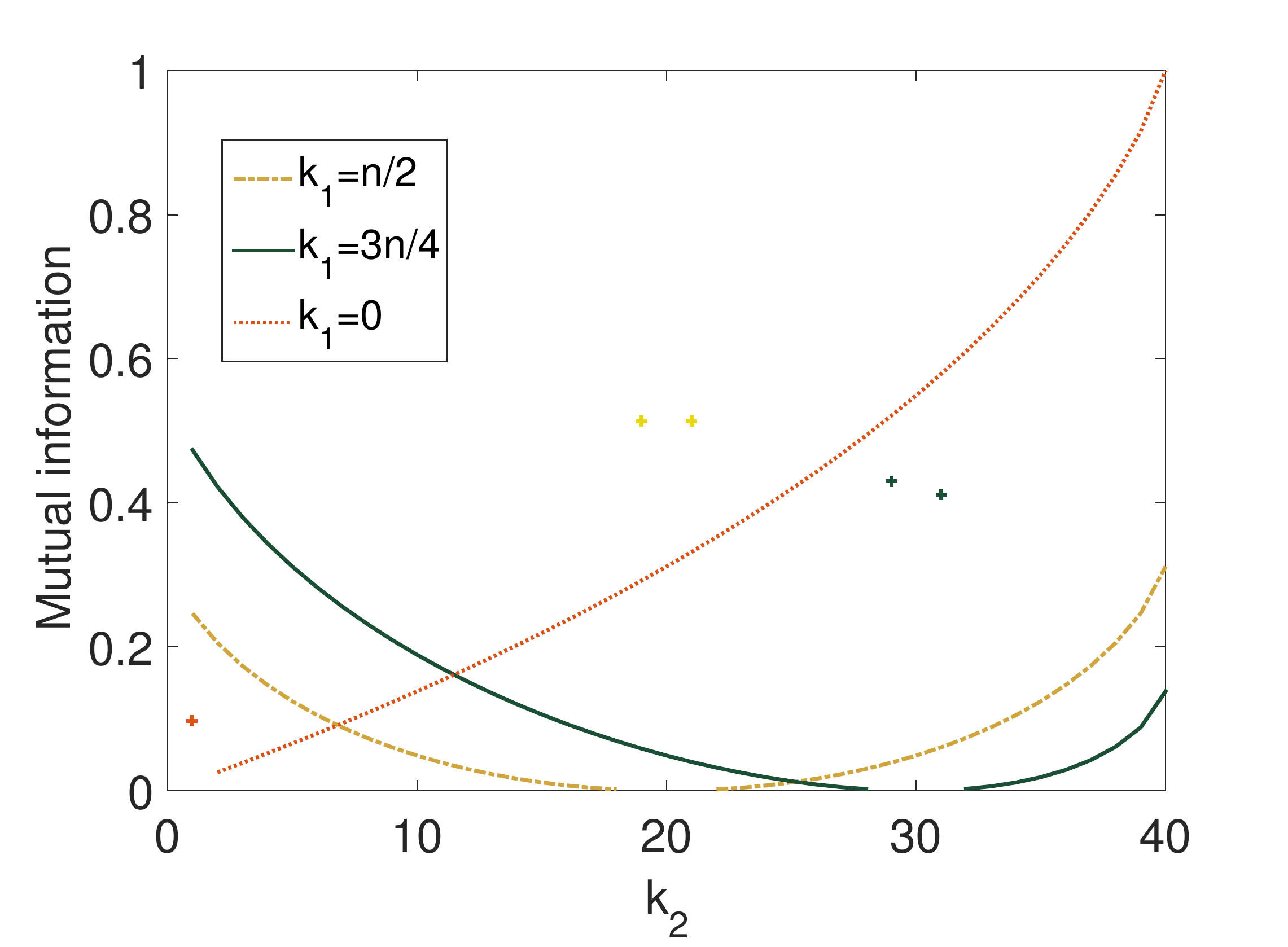}
\caption{\label{fig:mutualDicke}Mutual information for one party and different Dicke state encoding configuration. Single points represents isolated fluctuations when the difference on the number of excitation between codewords is one.}
\end{figure}

Following Sec. \ref{security}, we study the amount of information a single party can access once the quantum information has been delocalized over the network. The results are shown in figure \ref{fig:mutualDicke}. One finds again a relation with the local entanglement. We see that the entanglement is a key property that defines how the information is spread around the system, i.e. the less local entanglement, the more distributed. Hence, a single particle within a encoding configuration with low local entanglement has a low information about the global state of the logical system.

\subsection{Correlation space encodings}\label{correlationencoding}
In this section, we analyze the period wires (\ref{eq:matrixmps}) conceived as logical blocks, where the state of the correlation space defines the state of the logical qubit (see Sec. \ref{subsec:Correlation-space}).

\subsubsection{Uploading}
The general uploading procedure of Sec. \ref{subsec:Uploading} is
applicable for our wire configuration. In this case, a wire plays the role of the logical block, with codewords $\left|0_{L}\right\rangle=\Phi\left(\left|0\right\rangle \right) ,\left|1_{L}\right\rangle=\Phi\left(\left|1\right\rangle \right)$. Note that orthogonality of correlation space states does not always imply orthogonality of the corresponding wire states \cite{MorimaeUpload}. However, orthogonality is fulfilled for the period wires (\ref{eq:matrixmps}) we make use in this paper.

An alternative interpretation of uploading is possible \cite{MorimaeUpload}, where any physical state is uploaded into the correlation space. We review this process in the Appendix B.

\subsubsection{Downloading} \label{sec:downcorr}
In contrast to the Dicke-type encoding, it is known how to localize information of the correlation space into some physical site of a wire \cite{downloadDur}. This process does not imply the measurement of all the parties of the wire and we show that it can be made quasi deterministic. Based on the protocol developed in \cite{downloadDur}, we analyze how information is delocalized within the correlation space, and which properties influence this spreading of information.

We briefly review the downloading procedure of \cite{downloadDur}, and point out some additional features.  Given a quantum wire of the form (\ref{eq:matrixmps}),
one can always find a basis such that
\begin{equation}
A[m_{0}]=r_{0}\left|\varphi_{0}\right\rangle \left\langle 0\right|,\,A[m_{1}]=r_{1}\left|\varphi_{0}\right\rangle \left\langle 0\right|+\left|\varphi_{1}\right\rangle \left\langle 1\right|,\label{eq: matrix2}
\end{equation}
with $r_{0}>0$, $r_{1}\geq0$, $r_{0}^{2}+r_{1}^{2}=1$ and $\left\langle \varphi_{i}|\varphi_{j}\right\rangle =\delta_{ij}$.
In particular, if $r_{1}=0$ the operators are of the form:
\begin{equation}
A[m_{0}]=\left|\varphi_{0}\right\rangle \left\langle 0\right|,\,A[m_{1}]=\left|\varphi_{1}\right\rangle \left\langle 1\right|,\label{eq: matrix1}
\end{equation}
with $\left|\varphi_{0}\right\rangle =\cos\frac{\theta}{2}\left|0\right\rangle +e^{i\alpha}\sin\frac{\theta}{2}\left|1\right\rangle$
and $\left|\varphi_{1}\right\rangle =\sin\frac{\theta}{2}\left|0\right\rangle -e^{i\alpha}\cos\frac{\theta}{2}\left|1\right\rangle.$
In fact, we show that if one can find a basis $\left\{ m_{0},m_{1}\right\}$
such that matrices of the form (\ref{eq: matrix1}) exists, the local entropy of
entanglement of any particle of the wire is maximal for any $\theta$
(see Appendix C). Therefore, we can relate the local entanglement of the wire
with the probabilistic behavior of the localization process (see below).

Consider the case of a wire which admits a basis where the matrices are of the form (\ref{eq: matrix1}). The wire is prepared in the state
$\Phi\left(\left|\psi\right\rangle \right)_{k}^{n}$ by appropriate
measurements of the first $k$ particles. As before, by rewriting
the state as a function of the basis state of the $k+1$ position
and mapping the state of the correlation space into $\left\{ \left|\varphi_{0}\right\rangle ,\left|\varphi_{1}\right\rangle \right\} \rightarrow\left\{ \left|+\right\rangle ,\left|-\right\rangle \right\} $,
we end up with a state of the form: $\underset{s=0,1}{\sum}Z_{m}^{s}\left|\psi_{m}\right\rangle _{k+1}\left|m_{s}\right\rangle _{k'+1}\Phi\left(\left|\psi_{s}\right\rangle \right)_{k'+2}^{n}$. In order to accomplish the mapping, around $\tau$ measurements are needed (up to by-products), where $\tau$ is the period factor of the wire.
With a measure on the site $k'+1$ one achieves the downloading of
the state $\left|\psi\right\rangle $ at position $k+1$. For a more general wire configuration (\ref{eq: matrix2}), a filter operation \cite{downloadDur} is needed in order to recover orthogonality, leading to a probabilistic process for an a priori fixed site downloading.

In particular, we show that the success probability of localizing the information in the period wire (\ref{eq:matrixmps}) is directly related with its local entanglement.
The period wires are defined by the matrices (\ref{eq:matrixmps}), and the parameter $\phi$. As we discuss below (see Sec. \ref{locentcorr}) $\phi$ determines the local entanglement of the wire. For this wire, there exists a basis $\left|m_{i}\right\rangle$ such that the MPS matrices are expressed in the form (\ref{eq: matrix2}), with $r_{1}=\frac{1}{2}\left(1+e^{-i\phi}\right)$. Following \cite{downloadDur}, one concludes that the success probability of an a-priori fixed site downloading is $p=1-\left|r_{1}\right|=1-\cos\left(\frac{\phi}{2}\right)$. This shows that the probability depends exclusively on the entanglement factor, and the process is deterministic ($p=1$) for a maximally entangled wire $(\phi=\pi)$. Henceforth, the entanglement of the wire is one crucial property that defines how the information is spread over the particles.

If we do not restrict ourselves to the case where the localization has to be achieved in a specidfic, pre-defined site, and we allow for open destination downloading, the success probability can be made arbitrarily close to $1$ using larger wires. By repeating the filtering process $l$ times, the success probability becomes $p=1-\left|r_{1}\right|^{l}$. This is an important difference between the Dicke-type encoding and the correlation space,since here there is no need of measuring all the qubits to localize information and, therefore, if the localization procedure fails, one can keep repeating it until succeeding. Besides, once the process succeeds, information is localized in a qubit that is detached from the unmeasured part of the wire. The remaining part of the wire remains functional and can still be used to store and process information.

\subsubsection{Robustness under losses}
In an encoding scenario, each wire plays the role of a logical qubit
with codewords given by the states of the correlation space, typically
$\left|0_{L}\right\rangle=\Phi\left(\left|0\right\rangle \right) ,\left|1_{L}\right\rangle=\Phi\left(\left|1\right\rangle \right)$.

Consider again an auxiliary qubit entangled with a wire in the state
\begin{equation}
\left|\Phi^{+}\right\rangle =\frac{1}{\sqrt{2}}\left(\left|0\right\rangle _{aux}\Phi\left(\left|0\right\rangle \right)_{1}^{n}+\left|1\right\rangle _{aux}\Phi\left(\left|1\right\rangle \right)_{1}^{n}\right),\label{eq: maxent2}
\end{equation}
with $n\rightarrow\infty.$ Note this state can be obtained from
a single wire and a Bell pair, by performing a Bell measurement between one qubit of the Bell state, and the first particle of the wire, measuring the next $k$ particles to implement the rotation $\left\{ A[m_{0}]\left|R\right\rangle ,A[m_{1}]\left|R\right\rangle \right\} \rightarrow\left\{ \left|0\right\rangle ,\left|1\right\rangle \right\} $
in the correlation space.

\begin{figure}
\includegraphics[height=2.3in]{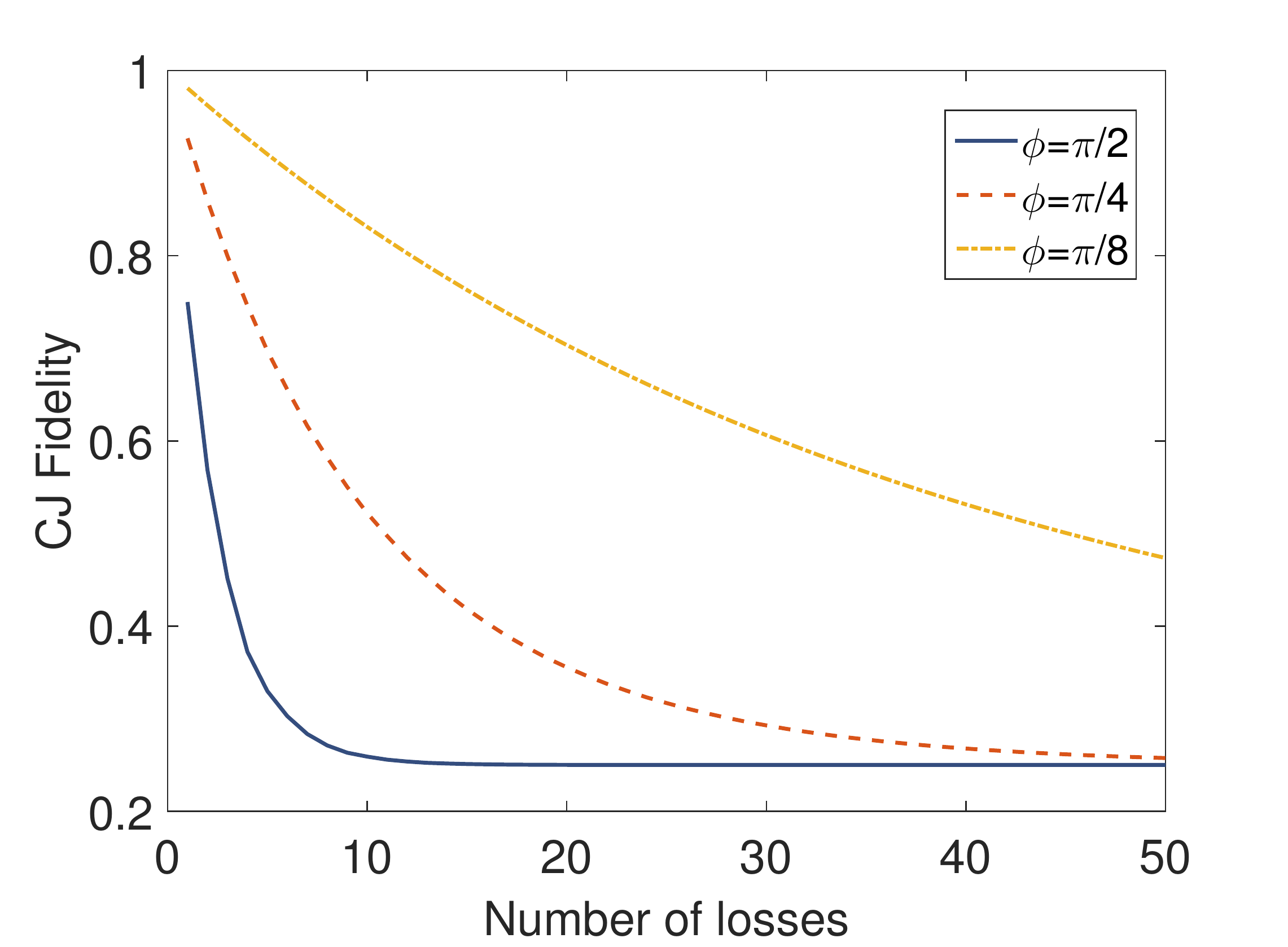}
\caption{\label{fig:cjfidelitycorrelation}CJ fidelity for different entanglement
factors of the wire. }
\end{figure}

We study the robustness
of this wire encoding under loss of particles, for different configurations.
For that purpose, we analyze again the CJ fidelity (\ref{eq:CJfidelity}) with respect
to the state (\ref{eq: maxent2}) when several particles of the wire
are lost (see figure \ref{fig:cjfidelitycorrelation}). Similarly to the Dicke-type encoding, the stability of this encoding under loss of particles depends on the local entanglement of the wire, where smaller entanglement leads to higher robustness. This confirms the relation between the entanglement and the spreading of information that we showed above.

\subsubsection{Local entanglement} \label{locentcorr}
\begin{figure}
\includegraphics[height=2.3in]{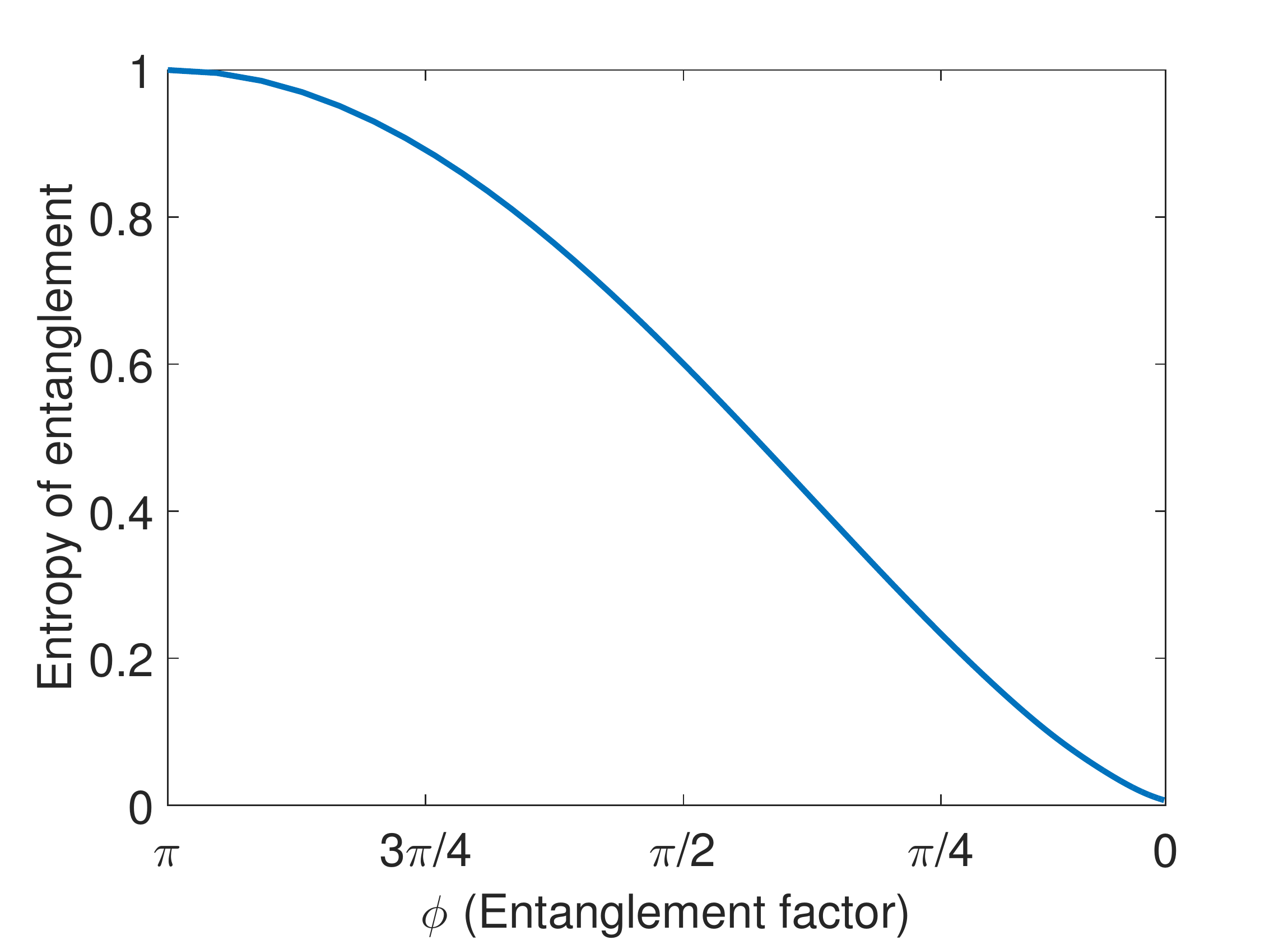}
\caption{\label{fig:entropycorrelation-1}Entropy of entanglement as a function of the entanglement factor of the wire.}
\end{figure}

In analogy to section \ref{subsec:Local-entanglement-(Entropy}, we
analyze the local entanglement of a single particle within a wire in the
logical state $\left|\Phi_{L}^{+}\right\rangle =\frac{1}{\sqrt{2}}\left(\left|0_{L}\right\rangle \left|0_{L}\right\rangle +\left|1_{L}\right\rangle \left|1_{L}\right\rangle \right)$, with $\left|0_{L}\right\rangle=\Phi\left(\left|0\right\rangle \right) ,\left|1_{L}\right\rangle=\Phi\left(\left|1\right\rangle \right)$
(see figure \ref{fig:entropycorrelation-1}). The parameter $\phi$ is the factor that determines the local entanglement of the wire, a result consistent with the studies of previous sections.  Therefore, denoting $\phi$ as entanglement factor is justified. Note (fig.  \ref{fig:entropycorrelation-1}) that the local entanglement of the wire can be made arbitrarily low.

\subsubsection{Security analysis}
\begin{figure}[ht!]
\includegraphics[height=2.3in]{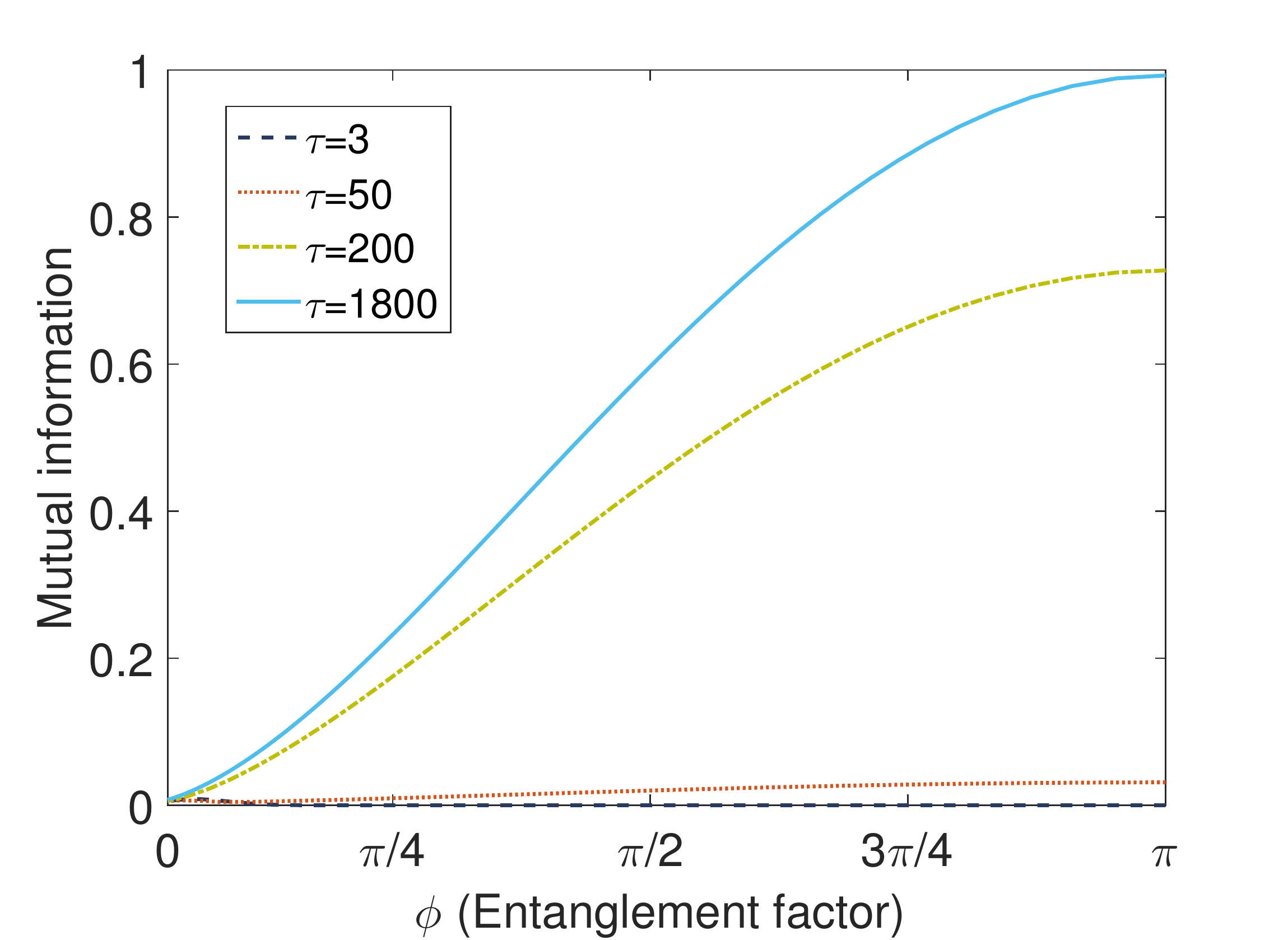}

\caption{Mutual information of one party and different wire period factor ($\tau$) configurations as a function of entanglement factor with $n=400$.}
\label{fig:mutualcorr}
\end{figure}

\begin{figure*}[ht!]
\subfloat{\includegraphics[height=2in]{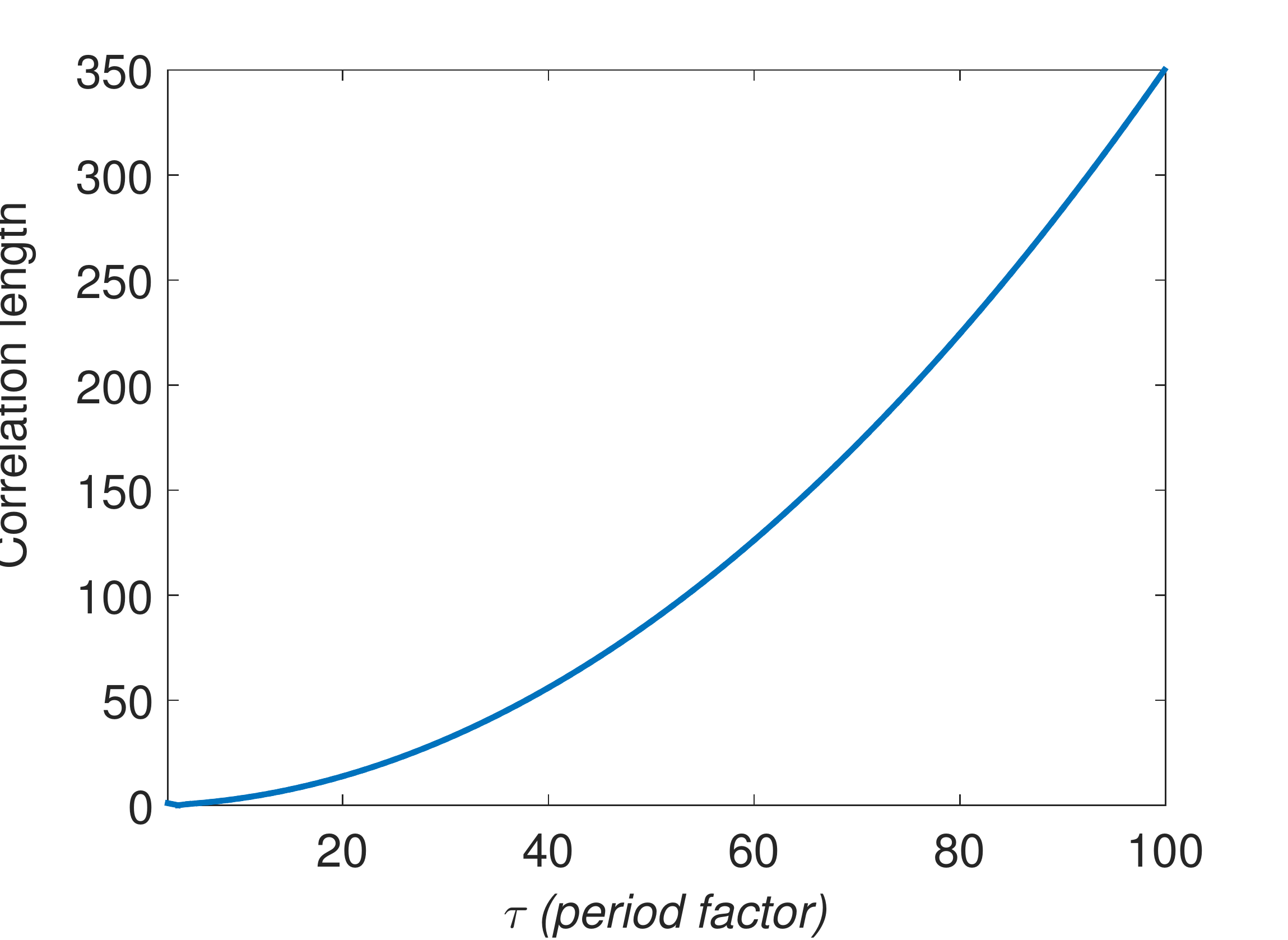}}
\subfloat{\includegraphics[height=2in]{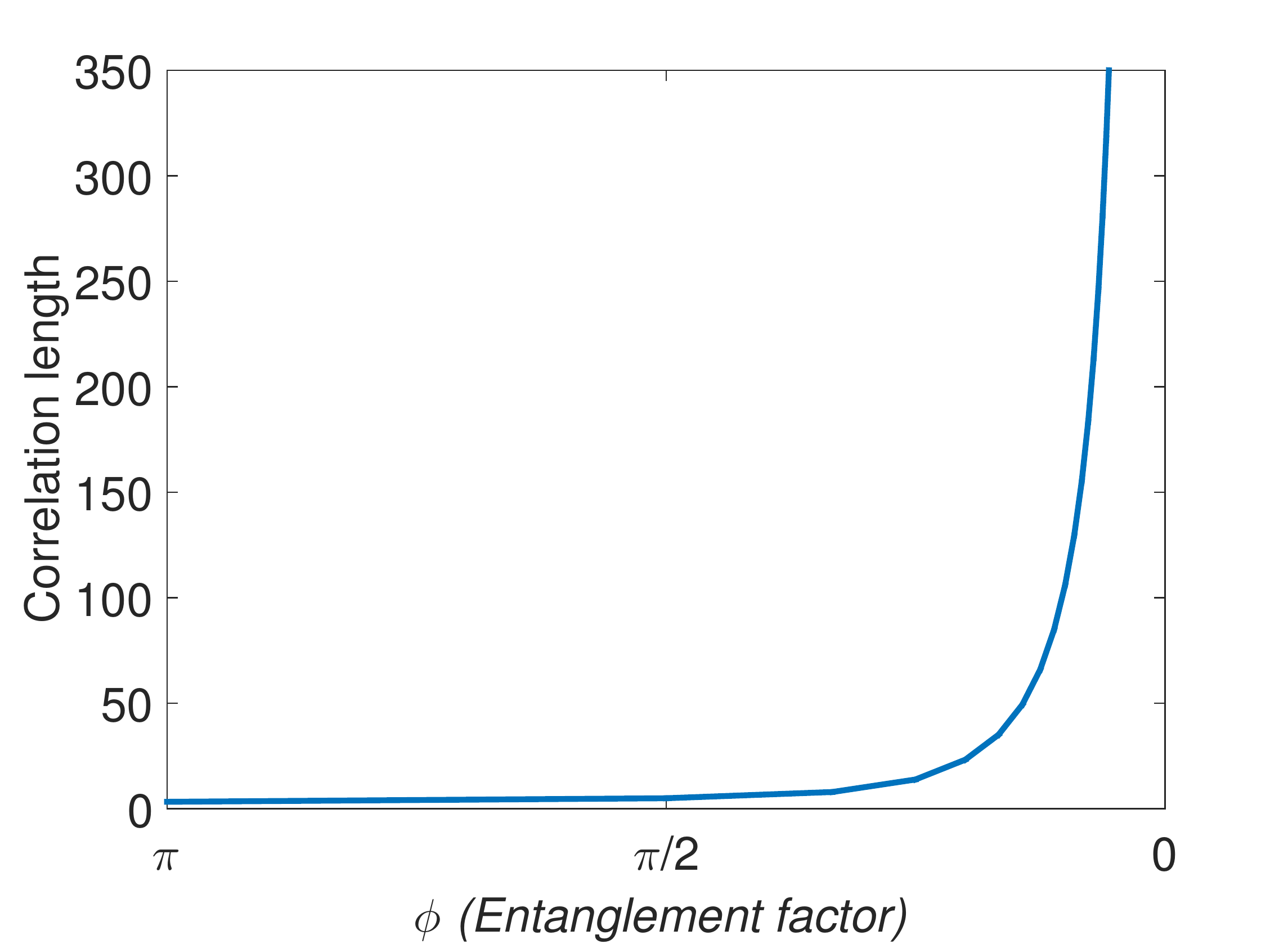}}
\caption{Correlation length of the period wire as a function of the period factor and the entanglement factor}
\label{fig:correleng}
\end{figure*}

We now analyze the amount of information available for a single party within the wire. Figure \ref{fig:mutualcorr} shows that the mutual information of a single party depends both on the entanglement and the period factors of the wire. In order to  understand these results we need to discern how the information of the correlation space is distributed throughout all the particles. The parameter we evaluate is the correlation length of the wire, that determines the distance over which information is spread in the wire. The correlation length is defined as \cite{mps2}:
\begin{equation}
\xi = \frac{-1}{\log\left(\frac{|\lambda_2|}{|\lambda_1|}\right)}, \label{correlenght}
\end{equation}
where $|\lambda_1|$ and $|\lambda_2|$ are the two largest eigenvalues of the transfer matrix $E_{\mathbb{I}}=\sum A[i]	\otimes \bar{A[i]}$. Figure \ref{fig:correleng} shows the dependence of the correlation length with the period and the entanglement factor separately. It is clear that both parameters have an important influence on the distribution of the information over the wire. This can be seen from the downloading process studied in previous sections. There we showed that the entanglement factor defines the success probability of the localization procedure. Moreover, in case one succeeds, a number of particles of the order of the period factor has to be measured in order to accomplish the localization.

We conclude that both, the local entanglement (entanglement factor) and the two-point correlation functions (period factor), are essential properties that determine how the information of the correlation space is distributed through the qubits of the wire. This is hence related to the amount of information accessible for each party, as we show in our analysis (fig. \ref{fig:mutualcorr}) i.e. the more distributed the information is, the less information a party can access. Besides, our analysis reveals that the position of the particle also matters. Particles close to the left boundary, where the state of the correlation space is defined (Sec. \ref{subsec:Correlation-space}), have a higher access to information than qubits sited away. The correlation length also determines how the amount of accessible information decreases with the distance to that left boundary.

\section{Quantum networks of encoded resources}\label{LogicalNetworks}
On can also build general entangled resource states that form a fully functional quantum network out of logical systems. To this aim, one considers an encoding $\{|0_L\rangle,|1_L\rangle\}$ and a multipartite resource state that is built form these logical qubits. Each logical qubit is formed by a region, i.e. a set of different parties, and the overall network consists of multiple of these logical qubits or regions. For instance, one may consider a multipartite logical GHZ state $|GHZ_L\rangle=|0_L\rangle^{\otimes n} + |1_L\rangle^{\otimes n}$, a logical 1D cluster state, or a general logical graph state network state \cite{Bruss1}. Each logical qubit serves to store one quantum bit of information, which is --similarly as discussed above-- distributed among the corresponding physical sites. One may then use such a resource to transport encoded quantum information, or to establish entanglement between logical qubits by measuring the other logical systems. For instance, one may generate from a logical GHZ state a logical Bell pair between any two logical qubits by measuring the remaining $(n-1)$ logical systems in the logical $X$-basis. A logical graph state can be manipulated by logical $Z$-measurements, which allows one to cut vertices from the graph, and by logical $X$ or $Y$-measurements, which allows one to transport quantum information among 1D structures \cite{Bruss2}, thereby realizing a general, entanglement based quantum network. Depending on the choice of logical resource state, one can also achieve full functionality in the sense that any desired logical graph state can be be generated between regions \cite{Pirker2,Bruss1,Bruss2,Markham2019}. For instance, a fully connected decorated logical graph state or a logical 2D-cluster state allow one to do this between all logical qubits, or a subset of them.

Interestingly, these logical measurements can always be done in a fully {\it local} way, that is by measuring only the individual physical qubits. This is a consequence of the result of \cite{walgate}, that states that any two orthorgonal multiqubit states can be deterministically distinguished by LOCC. The same procedure allows one to actually perform an arbitary projective measurement in the logical subspace, by chosing the eigenstates of the observable as the two states to be distinguished. Notice that no direct generalization of this result to distinguish $k$ orthogonal states exists (only a few special cases are known). Hence an extension of the protocol beyond logical qubits, i.e. to store and process logical $d$-level systems, cannot be directly obtained.

Notice that one may modify the scheme to distinguish between two orthorgonal states by LOCC to obtain a (probabilistic) download procedure. One selects the site to be downloaded as last qubit, and performs the LOCC protocol to distinguish between the two states as outlined in \cite{walgate}. Only the last step, i.e. the measurement of the last particle which determines the final outcome, is not performed. This performs a map $|0_L\rangle \rightarrow |\varphi_0\rangle$, $|1_L\rangle \rightarrow |\varphi_1\rangle$, where $\{|\varphi_k\rangle\}$ are orthogonal, but the states are not normalized. That is, the weights between the resulting states may change, depending on the branch of the LOCC protocol. This change of weights is in principle known, but needs to be undone to realize a full download of quantum information to physical sites. The final step corresponds to a filtering operation, which only succeeds probabilistically.



\section{Correlation space resources as a quantum network\label{sec:Correlationtransport}}
In this section we consider an alternative interpretation of the correlation
space of the quantum wires introduced before. Here, we understand
a quantum wire as a building block for a communication network, where each particle
represents one node of the network. We analyze the properties of the
wires when one demands processing of information (transport) and communication
between nodes (or regions) under request by only local operations, by exploiting the knowledge of the quantum wire of the previous sections.

\subsection{Period wire as a 1D quantum network}
Consider again the period quantum wire $\Phi$ of the form (\ref{eq: mpswire})
with $G=\exp\left(i\pi X/\tau\right)$ and $T=diag\left(e^{\frac{-i\phi}{2}},e^{\frac{i\phi}{2}}\right)$, where $\tau$ and $\phi$ are the period and the entanglement factor respectively.
The correlation space is prepared in some state $\left|\varphi\right\rangle $,
i.e. $\Phi\left(\left|\varphi\right\rangle \right)_{1}^{n}$. In order
to transport the information along the wire, one
just needs to perform $X$ measurements on the physical qubits. The
$X$ measurement implies the application of the $G$ operator in
the correlation space (up to by-products), such that the state is
rotated periodically throughout the virtual positions. The information
shows up completely
only every $\tau$ positions during transport. We analyze the influence
of errors during the transport by studying a processing-downloading
process.

\subsubsection{Errors during transport}
Consider an initial Bell state and a period quantum wire.
A generalized Bell measurement is performed between the second qubit of the Bell
pair and the first one of the wire. After by-product corrections (assuming
successful uploading), the remaining state is of the form:
\begin{equation}
\left|\Phi^{+}\right\rangle =\frac{1}{\sqrt{2}}\left(\left|0\right\rangle _{aux}\Phi\left(\left|0\right\rangle \right)_{k}^{n}+\left|1\right\rangle _{aux}\Phi\left(\left|1\right\rangle \right)_{k}^{n}\right).\label{eq:bell}
\end{equation}

We transport the state two periods ($2\tau$ positions) in the wire before starting a localization process.
In the ideal case (no error), the state of the correlation space $\Phi\left(\left|i\right\rangle \right)$
is first mapped into itself by the transport. Assuming the localization process is
successful (otherwise it could be repeated until succeeding), the
remaining state is again a Bell pair between the first qubit of the
initial Bell state (aux) and the qubit of the wire where downloading
succeeds, decoupled from the rest of the wire. In order to examine the stability of the wire under errors in the transport, we study the fidelity of this final state in case some error (or loss) occurs during the transport with respect to the ideal case (Bell pair obtained), with a similar spirit as for the CJ fidelity (\ref{eq:CJfidelity}).

When an error affects a physical particle, the corresponding error
on its matrix operator of the correlation space is in general non-trivial.
For instance, for a maximally entangled period wire with $\phi=\pi$,
if a Pauli error affect any particle, the effect on the
correlation space is (see Appendix D for proof):
\[
Z\left|j\right\rangle \longrightarrow A[j]Z,
\]
\[
X\left|j\right\rangle \longrightarrow XA[j]X,
\]
\begin{equation}
Y\left|j\right\rangle \longrightarrow iXA[j]ZX.\label{eq: logical error}
\end{equation}
Note that, when one transports information by measuring on the $X$
basis,
a $X$ error in the measured qubit has no influence on the correlation
space state.

For other entanglement configuration of the wire, these relations between the physical errors and they correspondence in the correlation space become cumbersome. However, if the qubit where any CPTP error occurs is measured in any basis (which is the case we typically come up against), it is known \cite{Morimae2012} which is the effect on the correlation space.

\begin{figure}
\includegraphics[height=2.3in]{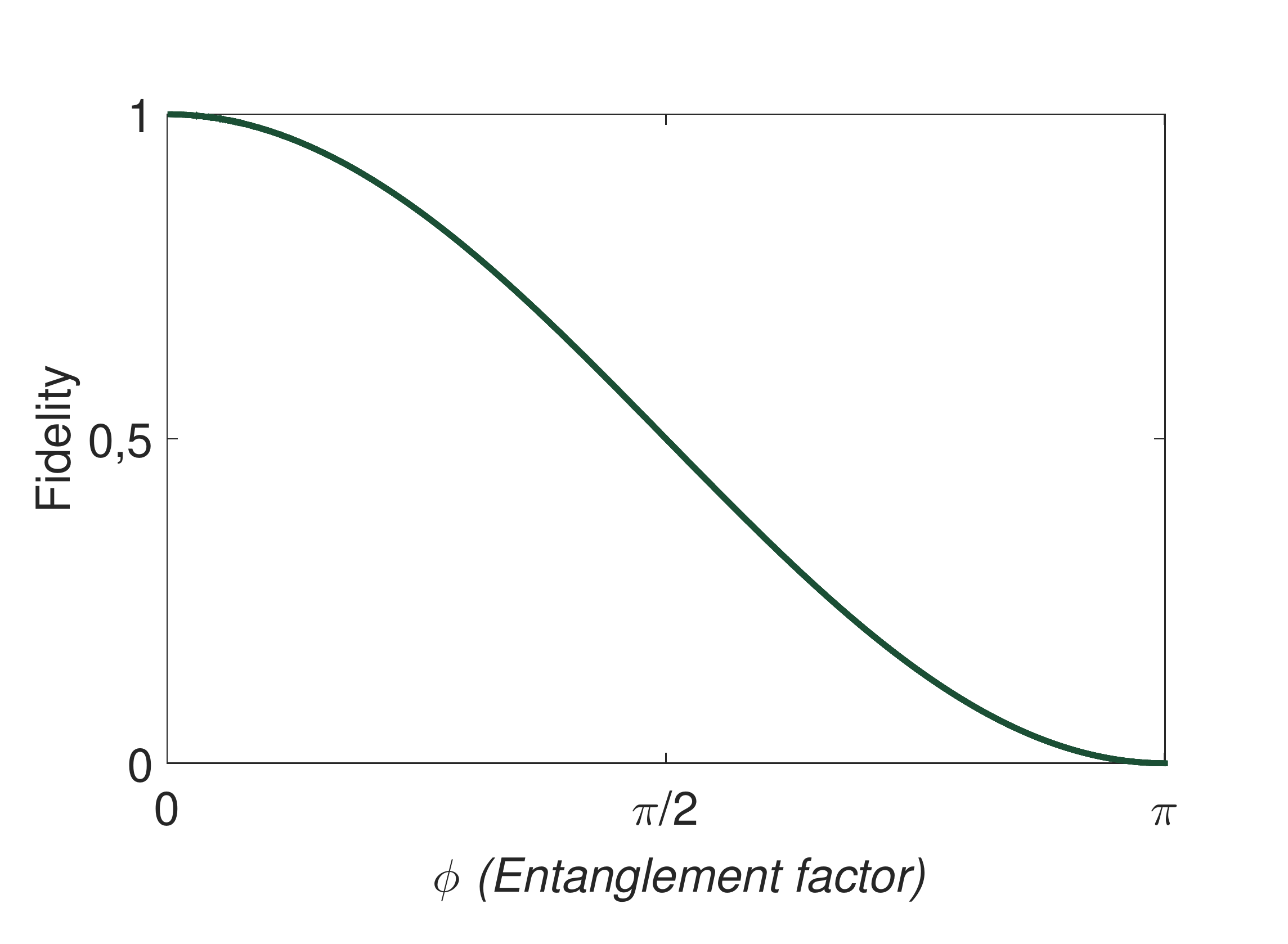}
\caption{\label{fig:fidelierror}Fidelity of the downloaded state with a Z
error occurring in the transport, as a function of the local entanglement
of the wire.}
\end{figure}

Consider now, for instance, a Pauli $Z$ error affecting the $k^{th}$
particle of the wire with probability one during a transport process. The state
$\Phi\left(\left|i\right\rangle \right)$ for $i=\left\{ 0,1\right\} $
is now mapped into some state $\Phi\left(\left|\chi_{i}\right\rangle \right)$
after the transport. Assume also that the subsequent downloading process
is successful. If the local entanglement of the wire is maximal, i.e.
$T=Z$, the error shows up completely in the end, i.e. the final state
is the Bell state $\left|\Phi^{-}\right\rangle =\frac{1}{\sqrt{2}}\left(\left|00\right\rangle -\left|11\right\rangle \right)$.
However, if one reduces the local entanglement of the wire (the parameter
$\phi$) , the error appears diluted at the end of the process. For
instance, for $\phi=\frac{\pi}{3}$, the final state expressed in
the Bell basis is: $\frac{\sqrt{3}}{2}\left|\Phi^{+}\right\rangle +\frac{1}{2}\left|\Phi^{-}\right\rangle $.
Figure \ref{fig:fidelierror} shows the fidelity of the final state
$\left|\varphi\right\rangle $ after the transport-localization process
with some $Z$ error affecting one particle in the transport, i.e.
$F=\left|\left\langle \varphi\left|\Phi^{+}\right\rangle \right.\right|^{2}$,
for different entanglement parameters. The remarkable result is
that this fidelity does not depend on the position where the error
happens or the period parameter of the wire. Therefore, the stability of the wire depends exclusively on the entanglement, as one can conclude due to the strong similarity
between figure \ref{fig:fidelierror} and figure \ref{fig:entropycorrelation-1}.

The price to pay for the protection against errors is a lower success
probability of delocalizing and localizing information.
However, when considering arbitrarily large wires, these processes can be made quasi-deterministic with open destination targets. One can always choose regions such that localization within this region has an arbitrary high probability of success. The size of this regions depends entirely on the local entanglement
of the wire. Notice that for a given success propability $p$ for the downloading process, the required number of repetitions to succeed with a probability $1-\epsilon$ is given by $n=\log(\epsilon)/\log(1-p)$. 

\subsubsection{Losses during transport}
The worst case of the previous scenario occurs when the particles
are lost. We consider the overall process of uploading--processing--downloading, where we assume that
one or more particles are lost (and hence traced out) during the transport. We
study again the fidelity with respect to the ideal
Bell state. Note that tracing out a particle is equivalent to averaging over the possible outcomes of a measurement in a given basis. By observation, one
can easily check that, in our case, this map is analogous to a phase-flip
channel $(\sigma\rightarrow q\sigma+(1-q)Z\sigma Z)$ with probability
$q=\frac{1}{2}$ at the physical level. The Kraus decomposition of this map can be
obtained by first expressing the final state in the Bell basis followed by a
subsequent diagonalization
\begin{multline}
\left|\psi_{00}\right\rangle \left\langle \psi_{00}\right|\overset{\xi}{\longrightarrow}\sum\lambda_{i_{1},j_{1},i_{2},j_{2}}\left|\psi_{i_{1}j_{1}}\right\rangle \left\langle \psi_{i_{2}j_{2}}\right|\overset{diag.}{\longrightarrow} \\
\overset{diag.}{\longrightarrow}\sum_{i}m_{i}\left|m_{i}\right\rangle \left\langle m_{i}\right|,\label{eq: kraus}
\end{multline}
where the Kraus operators $K_{i}$ are identified from: $K_{i}\left|\psi_{00}\right\rangle =\sqrt{m_{i}}\left|m_{i}\right\rangle $. For the case of one lost particle, this is always a rank-2 channel, independently
of the position of the lost qubit.

\begin{figure}
\includegraphics[height=2.3in]{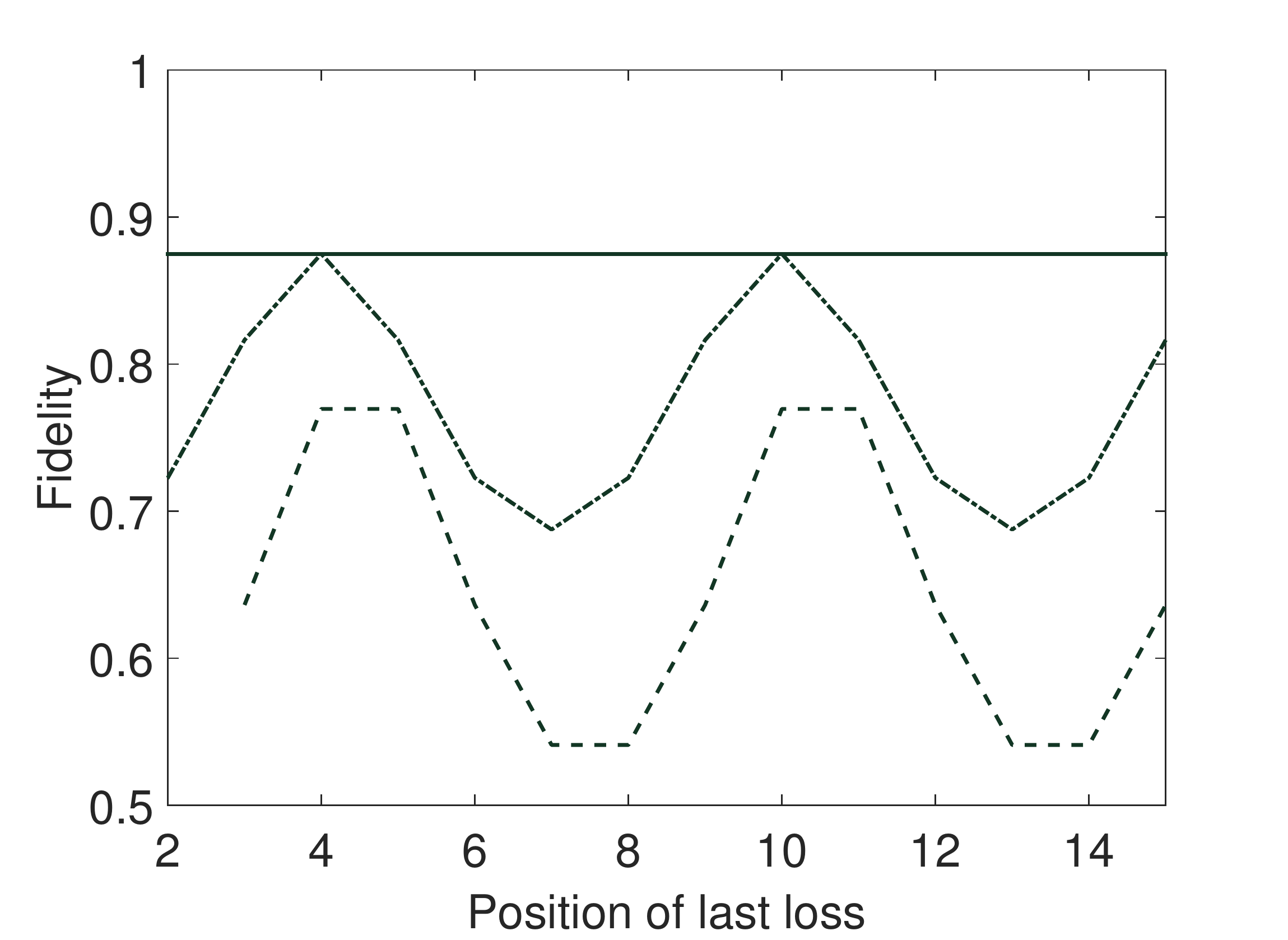}
\caption{\label{fig:losses} Dependence of the fidelity with the last lost particle.  Bold line corresponds to one single loss. Dashed-dot line represents two losses and the dependence with the second, where the first lost particles is fixed at site 1. Dashed line illustrates the case of three losses and the dependence with the third, where the first two lost particles are fixed in positions 1 and 2.}
\end{figure}

A remarkable effect is observed when more than one qubit is lost.
We find a fidelity dependence with the relative position of the lost
particles (see figure \ref{fig:losses}). The effect of the relative position
of the particles is strongly related with the period factor $\tau$ of
the wire. Given two losses, if the distance between them corresponds
with one period $\tau$, the fidelity is minimal. However, the fidelity
is maximized when the relative distance coincides with half of a period
$\frac{\tau}{2}$. In a more general scenario, with $m$ lost qubits,
one has to consider the $\tbinom{n}{m}$ relative distances of every
pair of lost particles. The fidelity is maximal (minimal) when the
maximal possible number of relative distances is equal to $\frac{\tau}{2}\left(\tau\right)$.
We conjecture that this effect is due to the non-vanishing two-point
correlation functions.

\subsection{Period wire as building block for general networks}
In this section we consider period wires as building blocks for general communication resources. We show that resource states for a fully functional communication network with arbitrary topology can be constructed by merging elementary wires in a suitable way. This then enables entanglement routing among the network. We start by listing the required functionality. One can consider a wire as a resource where quantum information can be uploaded, transported among the wire, and then downloaded at a chosen position. Similarly, one may also use such a wire to establish entanglement between two chosen sites. This corresponds to a double-downloading process as detailed below.

In turn, a network is given by multiple parties that are connected by wires. One may then upload information from a given site, download it to a chosen site as well as transport information among wires. Since in a general network a single node can be connected to several neighbours, one requires that the transport direction from a given node can be freely chosen. This includes the possibility to merge and cut wires, where the merging is only needed to establish resources \cite{Bruss1,Bruss2}.
Here we mainly consider point-to-point communication, i.e. using the network in such a way that information is transported from $A$ to $B$. One may also establish entanglement between the two chosen sites. We also show how certain graph states, including GHZ states, can be obtained in the network.

We show in the following how all elementary building blocks and required processes can be realized with correlation space resources.


It was shown in \cite{MorimaeUpload} how to upload some arbitrary state (placed in the zeroth site) into the correlation space by LOCC and local measurements. With analogous techniques used in \cite{MorimaeUpload} (see Appendix), it is straightforward to see that a state can be uploaded from any node of the wire, by choosing the direction of uploading within the wire. This involves cutting the wire in the opposite direction. The uploaded state can be subsequently transported to other regions in a protected way, and localized in another node.

\subsubsection{Cutting of resources}
It is possible to cut a chain from an arbitrary site of the wire in
any direction, by using similar techniques than for
localization (Sec.\ref{sec:downcorr}). For simplicity, consider a wire with maximal local entanglement:
\begin{equation}
\Phi\left(\left|L\right\rangle \right)_{1}^{n}=\sum\left\langle s_{n}\right|A\left[s_{n-1}\right]\cdots A\left[s_{1}\right]\left|L\right\rangle \left|s_{1}\ldots s_{n}\right\rangle .\label{eq:cut1}
\end{equation}
Taking property (\ref{eq: matrix1}) into account, one can expand
the $k^{th}$ site in its basis, i.e.
\begin{multline}
\Phi=\sum\left\langle s_{n}\right|A\left[s_{n-1}\right]\cdots\left|\varphi_{i}\right\rangle _{k}\left\langle i\right|\cdots A\left[s_{1}\right]\left|L\right\rangle \\
\left|s_{1}\ldots s_{n}\right\rangle \left|m_{i}\right\rangle _{k}.\label{eq:cut2}
\end{multline}
By appropriate measurements of the next $q$ sites, one can perform
the transformation $\left\{ \left|\varphi_{0}\right\rangle ,\left|\varphi_{1}\right\rangle \rightarrow\left|+\right\rangle ,\left|-\right\rangle \right\} $
in the correlation space. Expanding the site $r=k+q+1$ in its basis:
\begin{multline}
\Phi=\sum\left\langle s_{n}\right|A\left[s_{n-1}\right]\cdots\left|\varphi_{j}\right\rangle _{r\,k}\left\langle i\right|\cdots A\left[s_{1}\right]\left|L\right\rangle \\
\left|s_{1}\ldots s_{n}\right\rangle \left|m_{i}\right\rangle _{k}\left|m_{j}\right\rangle _{r}.\label{eq:cut3}
\end{multline}
Finally, by measuring site $r$ and selecting the outcome $\left|m_{0}\right\rangle _{r}$,
\begin{multline}
\Phi=\sum\left\langle s_{n}\right|A\left[s_{n-1}\right]\cdots\left|\varphi_{0}\right\rangle _{r\,k}\left\langle i\right|\cdots A\left[s_{1}\right]\left|L\right\rangle \\
\left|s_{1}\ldots s_{n}\right\rangle \left|m_{i}\right\rangle _{k}=\Phi\left(\left|\varphi_{0}\right\rangle \right)_{r'}^{n}\otimes\Phi\left(\left|L\right\rangle \right)_{1}^{k},\label{eq:cut4}
\end{multline}
one succeeds in cutting the chain into two wires, both preserving
their functionality, but separately. Note that the information of the
initial wire remains completely in the right one. In case the entanglement
is not maximal, one cannot select the position where the cutting is
accomplished, but it eventually succeeds, similar as in the downloading process.

\subsubsection{Double localization of information}
\begin{figure}
\includegraphics{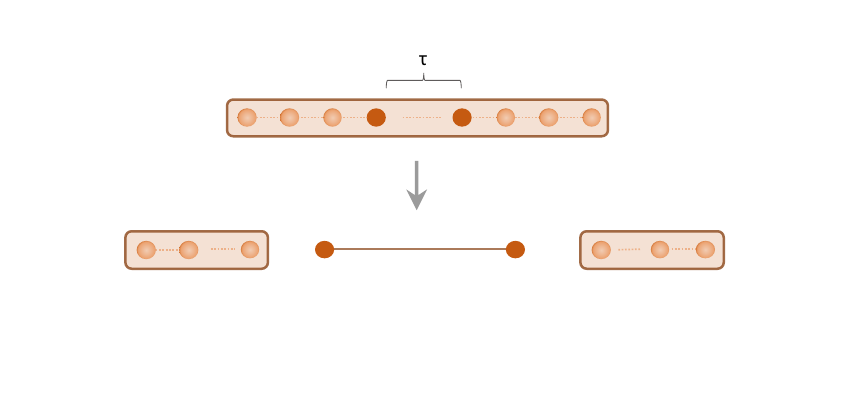}
\caption{\label{fig:Double-downloading}Localization of a Bell pair within a wire. Information is localized in two sites by processing a procedure similar to downloading along the external directions from both qubits.}
\end{figure}

Given a network constructed from period wires (see below), it is important to be able to connect
particles upon request, i.e. to establish entanglement between two (or more) parties. Consider again a maximally entangled
wire, i.e. $\phi=\pi$. Given two, previously fixed, qubits of the
wire, denoted as $k_{1}$ and $k_{2}$,
\begin{multline}
\Phi\left(\left|L\right\rangle \right)_{1}^{n}=\sum\left\langle s_{n}\right|A\left[s_{n-1}\right]\cdots\left|\varphi_{i}\right\rangle _{k_{1}}\left\langle i\right|\cdots\left|\varphi_{j}\right\rangle _{k_{2}}\left\langle j\right|\cdots \\
\cdots A\left[s_{1}\right]\left|L\right\rangle \left|m_{i}\right\rangle _{k_{1}}\left|m_{j}\right\rangle _{k_{2}}\left|s_{1}\ldots s_{n}\right\rangle,\label{eq:doubledow}
\end{multline}
one can create a direct communication link between them by applying
a double localization process (figure \ref{fig:Double-downloading}). This is achieved by measuring
the particles toward the exterior direction in each case, mapping
$\left|\varphi_{i}\right\rangle _{k_{1}}\rightarrow\left|i_{+}\right\rangle$
to the left, and $\left|j\right\rangle _{k_{2}}\rightarrow\frac{1}{\sqrt{2}}\left(\left|\varphi_{0}\right\rangle+\left(-1\right)^{j}\left|\varphi_{1}\right\rangle \right)$ to the right.
If there are $\tau-1$ particles between the two qubits, and we measure
the $\tau-1$ particles in the $X$ basis, where $\tau$ is the period factor of the wire,
a Bell pair is obtained between the selected qubits and the state is decoupled from
the rest of the wire in both directions (see figure \ref{fig:Double-downloading}).
In case the local entanglement of the wire is not maximal, this process becomes probabilistic and target sites cannot be fixed previously.

Note that this process has two inconveniences. First, not any pair of particles can be chosen, since one is restricted to the property of leaving one period between them. Secondly, in case that we obtain unwanted outcomes (by-products) in the measurement of the intermediate nodes, the final state is not a perfect Bell pair.

Nevertheless, both problems can be solved by considering compensation stations placed in different positions of the network. This stations are simply constructed with a certain number of auxiliary wires of limited length. These auxiliary wires are coupled with the main wires in order to process information until period restrictions and measurement by-products are fixed. In a sense, these boxes serve to extend the distance between two sites to a full period. All possibilities are present, and one only needs to cut all wires except the one with the proper length. Notice that the length of the wire can be properly chosen such that cutting and correcting suitable byproducts is possible.

\subsubsection{Merging of resources}
We conceive the 1D quantum wires as fundamental building blocks. In
order to construct more complex structures, one has to be able to
couple wires in order to route the information or create direct links
between any constituents.

\begin{figure}
\includegraphics{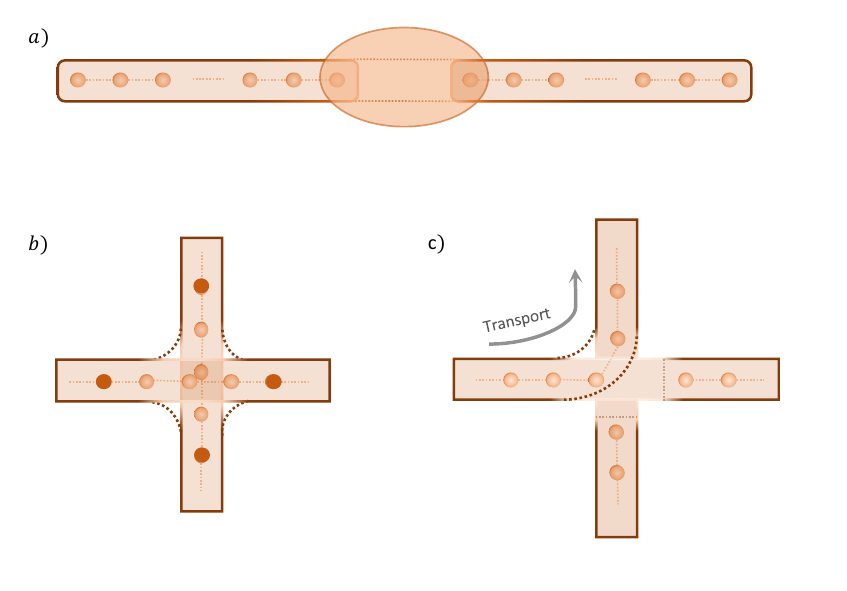}
\caption{\label{fig:merge}a) Two wires can be merged in their boundaries to create a larger wire. More wires can be merged in their boundaries to distill multipartite entangled states with one particle of each wire. b) Wires can be also merged in any middle point to localize n-party entangled states. c) By merging wires in any middle point, one can re-direct the transport of information of the correlation space along any other direction of the merged wires.}
\end{figure}

For simplicity, we restrict ourselves to the case of maximally entangled
wires. However, these processes are valid also
for lower entangled wires, by applying appropriate filtering operations
and taking into account the impossibility of a priori fixed destination.
We start by showing that two 1D wires can be coupled in the boundary qubits
to merge them into a larger wire. Given two wires initialized in the
$\left|+\right\rangle $ state,
\begin{multline}
\Phi\left(\left|+\right\rangle \right)_{1}^{n}\otimes\Phi'\left(\left|+\right\rangle \right)_{n'}^{k}=\sum\left\langle s_{k}\right|A\left[s_{k-1}\right]\cdots A\left[s_{n'}\right]\left|+\right\rangle \\
\left\langle s_{n}\right|A\left[s_{n-1}\right]\cdots A\left[s_{1}\right]\left|+\right\rangle \left|s_{1}\ldots s_{n}\right\rangle \left|s_{n'}\ldots s_{k}\right\rangle ,\label{eq:merge1}
\end{multline}
we perform the operation $P=\left|0\right\rangle \left\langle m_{0}0\right|+\left|1\right\rangle \left\langle m_{1}1\right|$
between the last qubit of the left chain and the first qubit of the
right one, merging them into an intermediate node $r$. The basis
$\left\{ m_{0},m_{1}\right\} $ corresponds to the one where operators
are expressed as eq. (\ref{eq: matrix1}). Note that for the wires
we use this basis coincides with the computational basis. Therefore,
one obtains
\begin{multline}
\Phi=\sum\left\langle s_{k}\right|A\left[s_{k-1}\right]\cdots A\left[i\right]\left|+\right\rangle \left\langle i\right|A\left[s_{n-1}\right]
\cdots A\left[s_{1}\right]\left|+\right\rangle \\
\left|s_{1}\ldots s_{n-1}\right\rangle \left|s_{n'+1}\ldots s_{k}\right\rangle \left|i\right\rangle _{r}= \\
=\sum\left\langle s_{k}\right|A\left[s_{k-1}\right]\cdots\left|\varphi_{i}\right\rangle \left\langle i\right|A\left[s_{n-1}\right]\cdots A\left[s_{1}\right]\left|+\right\rangle \\
\left|s_{1}\ldots s_{n-1}\right\rangle \left|i\right\rangle _{r}\left|s_{n'+1}\ldots s_{k}\right\rangle =\Phi\left(\left|+\right\rangle \right)_{1}^{k},\label{eq:merge2}
\end{multline}
and the merging successes (see figure \ref{fig:merge}a).

More than two wires can be coupled. One can consider, for instance,
three wires where a projection $P=\left|0\right\rangle \left\langle 000\right|+\left|1\right\rangle \left\langle 111\right|$
is performed on their extremities (first site of each), i.e.
\begin{multline}
\Phi=\sum\left\langle s_{n''}\right|A\left[s_{n''-1}\right]\cdots A\left[i\right]\left|+\right\rangle \\
\left\langle s_{n'}\right|A\left[s_{n'-1}\right]\cdots A\left[i\right]\left|+\right\rangle \left\langle s_{n}\right|A\left[s_{n-1}\right]\cdots A\left[i\right]\left|+\right\rangle \\
\left|s_{2}\ldots s_{n}\right\rangle \left|s_{2'}\ldots s_{n'}\right\rangle \left|s_{2''}\ldots s_{n''}\right\rangle \left|i\right\rangle _{r}.\label{eq:merge3}
\end{multline}
If now the states of the correlation space are all mapped $\left|\varphi_{i}\right\rangle \rightarrow\left|i\right\rangle $
by suitable local measurements and one expands the next site (denoted
as $q$):
\begin{multline}
\Phi=\sum\left\langle s_{n''}\right|A\left[s_{n''-1}\right]\cdots A[j]\left|0\right\rangle \left\langle s_{n'}\right|A\left[s_{n'-1}\right]\cdots A\left[j\right]\left|0\right\rangle \\
\left\langle s_{n}\right|A\left[s_{n-1}\right]\cdots A\left[j\right]\left|0\right\rangle \left|s_{q-1}\ldots s_{n}\right\rangle \\
\left|s_{q'-1}\ldots s{}_{n'}\right\rangle \left|s_{q''-1}\ldots s{}_{n''}\right\rangle \left|jjj\right\rangle _{qq'q''}\left|j\right\rangle _{r}.\label{eq:merge4}
\end{multline}

If now localization on each wire is performed (see previous section), by mapping the correlation
state to $\left|\pm\right\rangle $ and measuring the following site
in the computational basis, a 4-party GHZ state $\sum\left|jjj\right\rangle _{qq'q''}\left|j\right\rangle _{r}$
is obtained. By measuring the $r$ qubit in the X basis, a GHZ of
three qubits (one of each wire) is found and decoupled from the rest
of the wires (figure \ref{fig:merge}b). Note that the remaining wires
are still functional. Note also that by constructing router stations with auxiliary wires, any particle can be chosen on
demand to be part of this GHZ state. This process can be expanded for an arbitrary large number
of wires, therefore generating $n$-party GHZ states.

With analogous techniques, one can prove that two wires can be coupled
in any middle point of them, such that a four-party GHZ state can
be obtained by downloading information in the four diverging directions
(see figure \ref{fig:merge}b). Given this merging configuration,
one can easily check that a cutting can be performed in any of the directions, such that one can transport information through the remaining chain leafs (see figure \ref{fig:merge}c). That is, a network with a given structure (intersecting wires), one can either generate GHZ states between parties next to an intersection vertex, or use any of the intersecting vertices to route information. To this aim all but an incoming and outgoing wire are cut.

We emphasize that the merging of wires is a not process that takes place when processing information in the network. This should rather be understood as a tool that describes the required resource states in such an entanglement based network, and how they can (in principle) be generated from individual wire states.

\subsubsection{Fully functional communication network}
\begin{figure}
\includegraphics{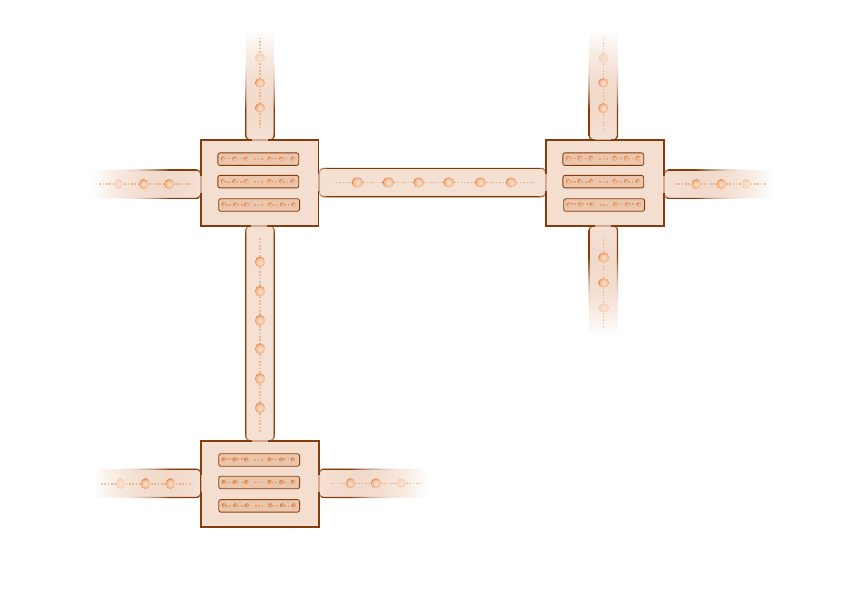}
\caption{\label{fig:wirenetwork} Construction example of a fully functional communication network based on period wires. Information can be protected during transport and direct links (multipartite states) can be obtained between constituents of any region with the help of the auxiliary wires in the compensation stations.}
\end{figure}

Given the quantum period wires and the tools studied above, it is possible to construct a networks with arbitrary topology where information can be transport through in a protected way, and where direct communication links between any two (or more) particles can be established. These is accomplished by merging 1D wires and including compensation stations between regions as described above (see fig. \ref{fig:wirenetwork}).


\section{Summary and Conclusions}
We have studied two different approaches for encoding and storage of quantum information in a distributed way. On the one hand, Dicke-type encodings (Sec. \ref{dickeenconding}) rely on an experimentally attainable class of states (Dicke states), where different orthogonal pairs of states can be selected exhibiting different properties. The measurement of all the parties is necessary to localize information, and the entanglement determines the success probability of this process. A conflict of interest exists when aiming to store information with stability under errors (or losses) and still having a high success probability to localize information.

On the other hand, we have analyzed the properties of the correlation space resource states, in particular of period wires (Sec. \ref{correlationencoding}) for storage of quantum information. We have seen how information is distributed over the wire. These wires are basically defined by two tunable parameters that define its entanglement and the correlation length. Therefore, localization of information (with LOCC) can be made quasi-deterministic by allowing for open-site destination, without compromising stability under errors or losses. Once the downloading succeeds, the qubit with the localized information is detached from the unmeasured wire, which remains functional. Thereby the probability of success for the download process depends on the entanglement factor, while the required number of particles to be measured to accomplish localization depends on the correlation length.

Moreover, we have shown that one can construct a fully functional communication network. This is either based on considering multipartite resource states that are comprised of logical qubits. In this case, information is distributed over regions that form the logical qubits, but can be processed and routed among the network solely by local operations on the individual sites.

Our second approach consists in using period wires as building blocks. 
We show how to cut and merge wires, and we also discuss how to upload, transport and localize information within the wires. By suitably coupling wires beforehand to form suitable resource states and by establishing compensation stations to compensate by-products, a 2D network can be built such that direct communication links can be obtained between any group of constituents from different regions of the network. This is done in such a way that information is distributed throughout the correlation space network in a diluted way, such that it is protected from errors or even loss of particles. In addition, information processing takes place solely by measuring individual (physical) qubits.
Entanglement and correlations of the wire determine how the information is distributed. The more spread the information is, the more protected it is. However, when information is widely distributed through the network, localization and cutting processes becomes probabilistic, and sometimes the target nodes cannot be fixed a priori. Therefore, we consider these networks particularly useful for communication between regions (instead of single nodes) with a high robustness under errors or attacks.

With our approach, we have shown an alternative view on quantum networks. The essential element is that information is stored in a distributed (or holographic) way within parts or the whole network, thereby protecting it against errors and losses while keeping the processing of information simple, i.e. using only measurements on individual sites. It would be interesting to extend our approach to the storage of multiple qubits in the network in a delocalized way. While it is straightforward to consider generalizations of the resources to store higher dimensional systems or multiple qubits, it is not clear if and how uploading, downloading and information processing can be achieved using only local operations on individual sites.

\section*{Acknowledgements}
This work was supported by the Austrian Science Fund (FWF) through project P30937-N27. W.D. also acknowledges support by the Erwin Schrödinger International Institute for Mathematics and Physics (ESI), where part of this work was completed.

\bibliographystyle{apsrev4-1}
\bibliography{holographNetw}

\appendix

\section{Concatenated encoding codewords}\label{apConcat}
Alternative extensions of Dicke-type encodings can be explored. For instance, one can consider
the concatenation of two Dicke states, with a similar spirit of the
concatenated-GHZ \cite{cGHZ}. By taking $\left|\tilde{0}\right\rangle =\left|n,k_{1}\right\rangle, \left|\tilde{1}\right\rangle =\left|n,k_{2}\right\rangle $,
one can define the following codewords:
\footnotesize
\[
\left|0_{L}\right\rangle =\frac{1}{\sqrt{2}}\left\{ \left[\frac{1}{\sqrt{2}}\left(\left|\tilde{0}\right\rangle ^{m}+\left|\tilde{1}\right\rangle ^{m}\right)\right]^{\otimes N}+\left[\frac{1}{\sqrt{2}}\left(\left|\tilde{0}\right\rangle ^{m}-\left|\tilde{1}\right\rangle ^{m}\right)\right]^{\otimes N}\right\} ,
\]
\begin{equation}
\left|1_{L}\right\rangle =\frac{1}{\sqrt{2}}\left\{ \left[\frac{1}{\sqrt{2}}\left(\left|\tilde{0}\right\rangle ^{m}+\left|\tilde{1}\right\rangle ^{m}\right)\right]^{\otimes N}-\left[\frac{1}{\sqrt{2}}\left(\left|\tilde{0}\right\rangle ^{m}-\left|\tilde{1}\right\rangle ^{m}\right)\right]^{\otimes N}\right\}, \label{eq:concatDicke2}
\end{equation}
\normalsize
where the particles are organized in $N$ groups of $m$ particles
each. Without going into details, our results indicate that this concatenated-Dicke
states can be a suitable intermediate choice for storing information.
Given two Dicke states codewords $\left|\tilde{0}\right\rangle ,\left|\tilde{1}\right\rangle$,
direct encoding brings certain values for the robustness and local entanglement
(see previous sections). However, if one considers their concatenated form (\ref{eq:concatDicke2}), the robustness is in general decreased and the local entanglement is increased to some extent, tuned
properties that, together with the block structure of the concatenated encoding, can be useful for some purposes.

\section{Uploading of information in correlation space}
It was shown in \cite{MorimaeUpload} how to upload some state into the correlation space by LOCC. Given an arbitrary state $\left|\psi\right\rangle _{0}=\lambda_{0}\left|0\right\rangle +\lambda_{1}\left|1\right\rangle $
(placed in the zeroth site) that we want to upload to the correlation
space, and given a wire $\Phi\left(\left|L\right\rangle \right)_{1}^{n}$,
one can write the initial state in terms of some orthogonal basis
$\left\{ \left|m_{s}\right\rangle \right\} $ at site $1$:
\begin{equation}
\Phi\left(\left|L\right\rangle \right)_{1}^{n}\otimes\left|\psi\right\rangle _{0}=\underset{i,s=0,1}{\sum}\lambda_{i}\Phi\left(A\left[m_{s}\right]\left|L\right\rangle \right)_{2}^{n}\otimes\left|m_{s}\right\rangle _{1}\otimes\left|i\right\rangle _{0}.\label{eq:mpsupload}
\end{equation}
A generalized Bell measurement is subsequently applied to complete
the uploading, i.e. by projecting onto $\left|B_{1}\right\rangle _{10}=\sum_{j}\left|m_{j}\right\rangle _{1}\otimes\left|j\right\rangle _{0}$
the remaining state of the wire is $\sum_{j}\Phi\left(\lambda_{j}A\left[m_{j}\right]\left|L\right\rangle \right)_{2}^{n}$.
Finally, by adequate local measurements on the next physical sites,
one can implement the basis change $\left\{ A\left[m_{s}\right]\left|L\right\rangle \right\} \rightarrow\left\{ \left|s\right\rangle \right\} $
in the correlation space and the upload is successful. However, this
process is not always successful. In case the states $A\left[m_{0}\right]\left|L\right\rangle $
and $A\left[m_{1}\right]\left|L\right\rangle $ are not orthogonal,
one would need some probabilistic filtering operation (in analogy to Sec. \ref{sec:downcorr}).

\section{Maximally entangled wire}
\begin{figure}
\includegraphics{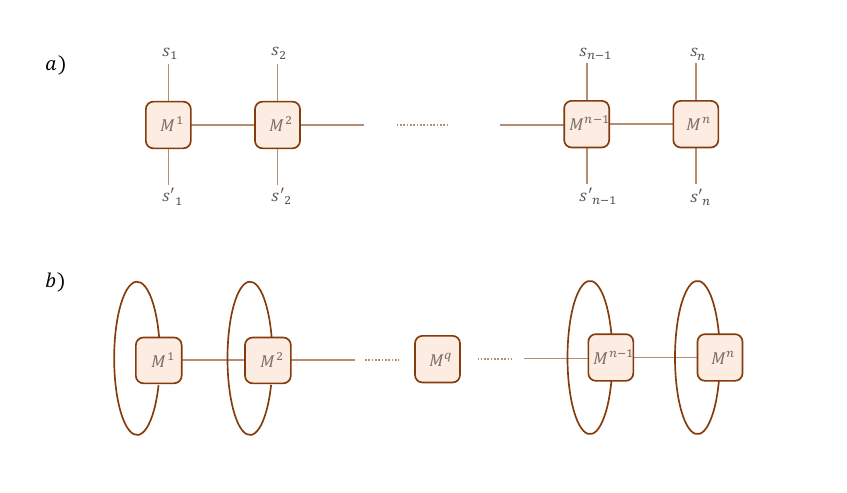}
\caption{\label{fig:Reduced-density-operator}a) Matrix product operator. b) Reduced density operator of a particle $q$, which is obtained by tracing out (contracting both physical inidices of each site) the rest of qubits.
MPO}
\end{figure}

We show that, given a quantum wire with period $k$ for which one
can find a basis $\left\{ m_{i}\right\} $ where $A[m_{0}]=\left|\varphi_{0}\right\rangle \left\langle 0\right|,\,A[m_{1}]=\left|\varphi_{1}\right\rangle \left\langle 1\right|$,
the local entanglement of any particle is maximal. We consider a general
case where $\left|\varphi_{0}\right\rangle =\cos\theta\left|0\right\rangle +\sin\theta\left|1\right\rangle $
and $\left|\varphi_{1}\right\rangle =\sin\theta\left|0\right\rangle -\cos\theta\left|1\right\rangle $,
for any $\theta$. In the period wires we use, the basis $\left\{ m_{i}\right\} $
coincides with the computational basis. Consider now a wire in the
state:
\begin{multline}
\Phi\left(\left|+\right\rangle \right)_{1}^{n}=\sum\left\langle s_{n}\right|A\left[s_{n-1}\right]\cdots A\left[s_{1}\right]\left|+\right\rangle \left|s_{1}\ldots s_{n}\right\rangle .
\end{multline}
One can compute the reduced density operator of an arbitrary particle
$q$, by first constructing the corresponding Matrix Product Operator
(MPO),
\[
\rho=\sum\left(M\left[s_{n},s'_{n}\right]\cdots M\left[s_{1},s'_{1}\right]\right)\left|s_{1}\ldots s_{n}\right\rangle \left\langle s'_{1}\ldots s'_{n}\right|
\]
with $D^{2}\otimes D^{2}$ operators $M\left[s_{i},s'_{i}\right]=\sum_{s,s'}A\left[s_{i}\right]\otimes\bar{A\left[s'_{i}\right]}$,
and subsequently tracing out the rest of particles. Note that tracing
out is equivalent to contracting indices $s_{i}$ and $s'_{i}$ in
every site $M$ (see figure \ref{fig:Reduced-density-operator}).
One can compute each element $\left(s_{q},s'_{q}\right)$ of the reduced
density matrix of site $q$ by contracting the expression from right
and left, i.e.
\[
\rho_{q}=\sum\left(M\left[s_{q-1}\right]M\left[s_{q},s'_{q}\right]M\left[s_{q+1}\right]\right)\left|s_{q}\right\rangle \left\langle s'_{q}\right|,
\]
where $M\left[s_{q-1}\right]$ and $M\left[s_{q+1}\right]$ are tensors
of dimension $1\otimes D^{2}$ and $D^{2}\otimes1$ respectively.
One can easily check that, from the right contracted boundary vector
$\left(\begin{array}{cccc}
1 & 0 & 0 & 1\end{array}\right)$, any new $1\otimes D^{2}$ vector carried to the left is again $E_{i}=\left(\begin{array}{cccc}
\sin^{2}\theta+\cos^{2}\theta & 0 & 0 & \sin^{2}\theta+\cos^{2}\theta\end{array}\right)=\left(\begin{array}{cccc}
1 & 0 & 0 & 1\end{array}\right),$ where $E_{i}$ is the intermediate transfer operator at position
$i$. With these techniques, it is straightforward to see that the
reduced density operator of the qubit $q$ (enough far away from boundaries)
is equal to the identity, independent of the value of $\theta$ and
the state of the correlation space, therefore proving that the entropy
of entanglement of the wire is maximal a local level, if a basis exists
such that $A[m_{0}]=\left|\varphi_{0}\right\rangle \left\langle 0\right|,\,A[m_{1}]=\left|\varphi_{1}\right\rangle \left\langle 1\right|.$

\section{Errors in the correlation space}
Given a maximally entangled period wire with $\phi=\pi$ and period
$k$, one can find a basis $\left\{ m_{i}\right\} $ where the correlation
space matrices can be expressed (\ref{eq: matrix1}) as $A[m_{0}]=\left|\varphi_{0}\right\rangle \left\langle 0\right|,\,A[m_{1}]=\left|\varphi_{1}\right\rangle \left\langle 1\right|.$
Noting that $\left\langle i\left|\varphi_{j}\right\rangle \right.=\frac{1}{2}\left(e^{w}+\left(-1\right)^{i\oplus j}e^{-w}\right)$
with $w=\frac{i\pi}{k}$, one can represent the wire state as:
\begin{multline}
\Phi\left(\left|+\right\rangle \right)_{1}^{n}=\sum\left\langle s_{n}\right|A\left[s_{n-1}\right]\cdots A\left[s_{1}\right]\left|+\right\rangle
\left|s_{1}\ldots s_{n}\right\rangle = \\
\sum\left(e^{w}+\left(-1\right)^{s_{n}\oplus s_{n-1}}e^{-w}\right)\cdots\left(e^{w}+\left(-1\right)^{s_{q}\oplus s_{q+1}}e^{-w}\right) \\
\left(e^{w}+\left(-1\right)^{s_{q}\oplus s_{q-1}}e^{-w}\right)\cdots\left(e^{w}+\left(-1\right)^{s_{1}\oplus s_{2}}e^{-w}\right) \\
\left|s_{1}\ldots s_{q}\ldots s_{n}\right\rangle .\label{eq:appendix1}
\end{multline}
If a Pauli $X$ error affects now the $q$ qubit, $X\left|s_{q}\right\rangle =\left|s_{q}\oplus1\right\rangle $.
Therefore, by relabelling in the equation (\ref{eq:appendix1}), the
state after the error is
\begin{multline}
\Phi=\sum\left(e^{w}+\left(-1\right)^{s_{n}\oplus s_{n-1}}e^{-w}\right)\cdots\left(e^{w}+\left(-1\right)^{s_{q}\oplus s_{q+1}\oplus1}e^{-w}\right) \\
\left(e^{w}+\left(-1\right)^{s_{q}\oplus s_{q-1}\oplus1}e^{-w}\right)\cdots\left(e^{w}+\left(-1\right)^{s_{1}\oplus s_{2}}e^{-w}\right) \\
\left|s_{1}\ldots s_{q}\ldots s_{n}\right\rangle .\label{eq:appendix2}
\end{multline}
It is easy to check that the effect on the correlation space is equivalent
to the map $A^{q}\left[m_{i}\right]\rightarrow XA^{q}[m_{i}]X=\left|\varphi_{i\oplus\text{1}}\right\rangle _{q}\left\langle i\oplus1\right|,$
on the virtual operator corresponding to the particle $q$. Equivalently,
if a Pauli $Z$ error applies to qubit $q$, i.e. $Z\left|s_{q}\right\rangle =\left(-1\right)^{q}\left|s_{q}\right\rangle $,
can be directly seen from the form of matrices $A[m_{0}]$ and $A[m_{1}]$,
that the translation of this error into the correlation space operator
is $A^{q}\left[m_{i}\right]\rightarrow A^{q}[m_{i}]Z$.

For Pauli $Y$ error, one just need to combine the previous cases
to find expression (\ref{eq: logical error}).

\end{document}